\documentclass[a4paper,11pt]{article}
\pdfoutput=1

\usepackage[utf8]{inputenc}
\usepackage{jheppub}
\usepackage{amsmath,amssymb,mathtools}
\usepackage{enumerate}
\usepackage{graphicx}
\usepackage{xspace}

\allowdisplaybreaks

\DeclareRobustCommand{\ensuremathrm}[1]{\ensuremath{\mathrm{#1}}\xspace}

\DeclareRobustCommand{\GeV}{\ensuremathrm{GeV}\xspace}
\DeclareRobustCommand{\TeV}{\ensuremathrm{TeV}\xspace}

\DeclareRobustCommand{\nnlojet}{NNLO\scalebox{0.8}{JET}\xspace}
\DeclareRobustCommand{\hybIso}{{\tt{hybIso}}\xspace}
\DeclareRobustCommand{\dynIso}{{\tt{dynIso}}\xspace}
\DeclareRobustCommand{\fixIso}{{\tt{fixIso}}\xspace}

\preprint{{\raggedleft%
  IPPP/19/23 \\
  ZU-TH 13/19 \\
  CERN-TH-2019-034 \\
}}

\title{Isolated photon and photon+jet production at NNLO QCD accuracy}

\author[a]{Xuan Chen,}
\author[a]{Thomas Gehrmann,}
\author[b]{Nigel Glover,}
\author[a]{Marius H\"ofer,}
\author[c]{Alexander Huss}
\affiliation[a]{Physik-Institut, Universit\"at Z\"urich, Winterthurerstrasse 190, CH-8057 Z\"urich, Switzerland}
\affiliation[b]{Institute for Particle Physics Phenomenology, Durham University, Durham, DH1 3LE, UK}
\affiliation[c]{Theoretical Physics Department, CERN, CH-1211 Geneva 23, Switzerland}
\emailAdd{xuan.chen@uzh.ch}
\emailAdd{thomas.gehrmann@uzh.ch}
\emailAdd{e.w.n.glover@durham.ac.uk}
\emailAdd{mhoefer@physik.uzh.ch}
\emailAdd{alexander.huss@cern.ch}

\abstract{%
We describe the calculation of the next-to-next-to-leading order (NNLO) QCD corrections to isolated photon and photon-plus-jet production, and discuss how the experimental hadron-level photon definition and isolation criteria can be approximated in the theoretical parton-level calculation. The NNLO corrections lead to a considerable reduction of the theory uncertainty on the predictions, typically to less than five per cent, and enable an improved description of experimental measurements from ATLAS and CMS\@.
}
\keywords{QCD, NNLO Computations, Hadronic Colliders, Jets}

\begin{document}
\maketitle


\section{Introduction}
\label{sec:intro}

Photon production at large transverse momenta is a classical hadron collider observable. It was measured already at early experiments 
at the ISR~\cite{Diakonou:1979sv,Anassontzis:1982gm,Angelis:1989zv} and Sp$\bar{\text{p}}$S~\cite{Albajar:1988im,Alitti:1991yk} colliders, 
followed by precision studies at the Tevatron~\cite{Aaltonen:2017swx,Abazov:2005wc} and by the ATLAS~\cite{Aad:2011tw,Aad:2016xcr,Aaboud:2017cbm}  
and CMS~\cite{Chatrchyan:2011ue,Sirunyan:2018gro} experiments at the LHC. In addition to these measurements of inclusive photon production, 
more exclusive photon-plus-jet final states were also investigated at the Tevatron~\cite{Abazov:2008er,D0:2013lra} and by ATLAS~\cite{Aad:2013gaa,Aaboud:2016sdm,Aaboud:2017kff} 
and CMS~\cite{Chatrchyan:2013oda,Chatrchyan:2013mwa,Khachatryan:2015ira,Sirunyan:2018gro}.

The underlying parton-level process~\cite{Halzen:1978rx,Ruckl:1978mx} is photon radiation off a quark in 
quark--antiquark annihilation or quark--gluon scattering, thereby providing~\cite{Harriman:1990hi,Vogelsang:1995bg,dEnterria:2012kvo,Carminati:2012mm,Campbell:2018wfu} 
sensitivity to the gluon distribution in the proton already at leading order. 
The measurement of photon production cross sections at hadron colliders and its interpretation is however more involved than 
it may appear at first sight, since besides this hard (``prompt'') radiation process
other processes may also yield final-state photons at large transverse momentum. In particular, photons can be radiated in an ordinary jet production event 
in the course of the hadronization process. This photon fragmentation process is described by (non-perturbative) fragmentation functions 
of different partons into photons~\cite{Koller:1978kq,Laermann:1982jr}. 
To minimize the contribution from the fragmentation process, the photon is required to be separated from any final-state hadrons in the 
event. This isolation requirement is typically formulated by admitting only a limited amount of hadronic energy inside a fixed-size cone around the photon 
direction. Lowering this energy threshold to zero is not possible for a finite-sized cone. This would restrict the phase space for soft parton 
emissions at higher order in QCD, thereby violating infrared safety of the definition of the observables.  Consequently, the theory description for a
fixed cone size must include contributions from photon fragmentation. 
An alternative isolation procedure uses a dynamical cone~\cite{Frixione:1998jh}, which lowers the hadronic energy cut towards the center of 
the isolation cone, thereby eliminating the fragmentation contribution. All experimental measurements to date use a fixed-size cone isolation.  

Photon production and photon-plus-jet production have the same underlying parton-level process, and differ only by 
the kinematical selection of the final state. 
Higher-order corrections to both processes turn out to be very sizeable. They have been computed to next-to-leading order (NLO) in QCD 
for a fixed-cone isolation procedure for inclusive photon~\cite{Aurenche:1987fs,Baer:1990ra,Aurenche:1992yc,Gordon:1993qc,Gluck:1994iz,Catani:2002ny}
and photon-plus-jet~\cite{Aurenche:2006vj} production. Using a dynamical cone isolation simplifies the theoretical description at 
higher orders, since fewer infrared-singular configurations need to be accounted for. With this isolation procedure, next-to-next-to-leading order (NNLO) 
QCD corrections were computed for both inclusive photon and photon-plus-jet production~\cite{Campbell:2016lzl,Campbell:2017dqk}. However, comparison 
of these predictions with data requires an empirical adjustment of the dynamical cone isolation parameters to mimic the effect of the fixed cone isolation 
used in the experimental measurements. 

In this paper, we present a new calculation of the NNLO QCD corrections to isolated photon and photon-plus-jet production. We use a combined isolation 
procedure, described in Section~\ref{sec:photondef}, which starts from the fixed-cone prescription, regulated by a considerably smaller dynamical cone, and 
quantify the parametric  uncertainties of this procedure. The calculation of the NNLO QCD corrections uses the antenna subtraction method and is performed in the 
\nnlojet framework. It is described in Section~\ref{sec:calc}. We present predictions for isolated photon and photon-plus-jet production in Sections~\ref{sec:isolated} 
and~\ref{sec:jet}, where they are compared with previous NNLO results and with recent data from ATLAS and CMS. We summarize our findings in Section~\ref{sec:conc}.

\section{Photon definition and isolation}%
\label{sec:photondef}

Photons that are produced at particle colliders are identified as deposits of electromagnetic energy. If an event under consideration 
contains highly energetic hadrons, their decay may mimic a photon signature (e.g.\ the decay $\pi^0\to \gamma\gamma$, if 
both photons are too collimated to be resolved individually). In order to separate photons produced 
in the hard scattering process from those produced through hadron decays, one commonly restricts the amount
of hadronic energy in the vicinity of a photon candidate, leading to an isolation criterion as part of the photon definition. 
A perfectly isolated photon would admit no hadronic energy in a certain region (typically a cone in pseudorapidity 
$\eta$ and azimuthal angle $\phi$) 
around the photon. This definition of isolation is however neither experimentally feasible nor theoretically well defined.
In order to construct an observable that is infrared safe under QCD corrections, emission of soft partons must be admitted 
everywhere in the final-state phase space. To prevent the photon isolation cone from obstructing the cancellation 
of infrared divergences between soft real radiation and virtual corrections, it must allow a finite amount of parton  
energy to be deposited close to the photon. Two types of cone-based isolations are being used, with $E_T^{{\rm had}}(R)$
denoting the sum of hadronic (partonic) energy inside a cone of  radius $R$ in the $(\eta,\phi)$-plane 
around the photon candidate direction:
\begin{enumerate}
    \item {\bf Fixed cone isolation:} 
    A cone with fixed radius $R$ is considered. If $E_T^{{\rm had}}(R)$ is smaller 
    than a threshold value, the photon candidate is considered to be isolated, and identified as a photon.
    $E_T^\mathrm{max}$ can be defined by a  fixed value, or as function of transverse energy $E_T^\gamma$ of the photon candidate:
    \begin{equation}
    E_T^{{\rm had}} (R) < E_T^\mathrm{max}\left(E_T^\gamma \right)\,,
    \end{equation}
    where for the purpose of this paper and following the experimental papers considered therein, we choose a simple linear dependence of $E_T^\mathrm{max}$ on $E_T^\gamma$, parametrized by
    \begin{equation}
    E_T^\mathrm{max} = \varepsilon E_T^\gamma + E_T^\mathrm{thres}\,.
    \end{equation}
     The fixed cone isolation procedure 
    can be implemented in a standard manner in the 
    experimental analysis, and is used in all experimental measurements of cross sections involving photons to date. When implemented in a theoretical calculation, however, it 
    introduces a sensitivity to the photon fragmentation process.  
    \item {\bf Dynamical cone isolation (Frixione isolation~\cite{Frixione:1998jh}):} 
    Starting from a cone with radius $R_d$, smaller concentric sub-cones with $r_d\leq R_d$ are considered. 
    The allowed hadronic energy 
    $E_T^{{\rm had}}(r_d)$ decreases with decreasing $r_d$, reaching zero for $r_d=0$. The photon candidate is accepted if the admitted hadronic energy is not exceeded 
    for any sub-cone radius $r_d$. The criterion is formulated using the functional form (with free parameters $\varepsilon_d$ and $n$):
    \begin{equation}
    E_T^{{\rm had}} (r_d) < \varepsilon_d E_T^\gamma \left(\frac{1-\cos r_d}{1-\cos R_d}\right)^n\, \quad \mbox{for all } r_d<R_d\,.
    \end{equation}
    With this, exactly collinear hard radiation in the photon direction is forbidden, while soft radiation is admitted over the whole phase space. Consequently, the theory prediction is 
    independent on the photon fragmentation process, which leads to a considerable simplification in its evaluation at higher orders. Because of the finite 
    detector resolution, the dynamical cone isolation can only be approximated in the experimental data analysis and it has not been used in any actual measurements to date. 
\end{enumerate}
Owing to the lower computational complexity associated with the dynamical cone isolation procedure, calculations  of NNLO QCD corrections to 
direct photon~\cite{Campbell:2016lzl,Campbell:2017dqk} and photon-pair production~\cite{Catani:2011qz,Campbell:2016yrh} have been performed using
this prescription only. To compare these calculations with measurements, which are all based on a fixed cone isolation procedure, the 
dynamical cone isolation parameters are adjusted~\cite{Balsiger:2018ezi} to mimic the effect of the fixed cone isolation. A detailed study of the effects of this approximation 
for photon pair production~\cite{Andersen:2014efa,Badger:2016bpw,Amoroso:2020lgh,Catani:2018krb} indicates its viability for sufficiently tight~\cite{Catani:2013oma} isolation criteria. 

The discrepancy between the isolation procedure used 
in experimental measurement and theory calculation is nevertheless unsatisfactory, and prevents quantitative statements on the impact of varying isolation parameters or predictions for 
loose photon isolation. Fully consistent NNLO predictions with a fixed cone isolation will require the computation of fragmentation contributions to this order, demanding 
an extension of NNLO methods towards identified final state particles. An improvement over the present predictions can however already be obtained by the following hybrid prescription~\cite{Siegert:2016bre}, 
which was used by the ATLAS collaboration in Ref.~\cite{Aaboud:2017kff} to compare data with NLO predictions from the multi-purpose SHERPA event generator~\cite{Gleisberg:2008ta,Hoeche:2012yf}:
\begin{enumerate}
    \setcounter{enumi}{2}
    \item {\bf Hybrid cone isolation:} 
    In the theoretical prediction, a 
    dynamical cone isolation with a small value of $R_d$ is combined with a fixed cone isolation with a larger value of $R$, such that $R^2\gg R_d^2$.
    The dynamical cone isolation is applied first, such that events very close to the collinear divergence are vetoed, and the dependence on the fragmentation process is 
    eliminated in an infrared-safe manner. The fixed isolation cone criterion is
    then applied to all events that passed the dynamical cone isolation. 
    The experimental analysis uses only the fixed cone isolation. Upon changes of the cone size $R$,
    the hybrid procedure exactly reproduces the behaviour of the fixed-cone isolation used in experiment: changing $R$ amounts to 
    modifying the catchment area used in the computation of   $E_T^{{\rm had}}(R)$.  The hybrid isolation procedure only 
    discards some events within a small inner fraction of the cone area, potentially introducing an unknown $R$-independent shift of the total amount of $E_T^{{\rm had}}(R)$. 
\end{enumerate}
In the following, we will perform a detailed comparison and parameter study of the effect of dynamical and hybrid cone isolation on the prediction of the photon-plus-jet production cross section.

\subsection{Comparison of isolation criteria and parameters}%
\label{subsec:ISO_param_dep}

The fixed-cone isolation  procedure admits a finite amount of collinear radiation in the photon direction, which induces a dependence on the 
photon fragmentation functions. Similar to parton distributions in the proton, the non-perturbative photon 
fragmentation functions~\cite{Koller:1978kq,Laermann:1982jr} fulfil  a QCD evolution equation whose boundary conditions can only be determined from experimental 
data. Up to now, only two experimental measurements of observables with direct sensitivity to the photon fragmentation function were performed, both in 
$e^+e^-$ collisions at LEP\@: 
inclusive photon production by the OPAL experiment~\cite{Ackerstaff:1997nha} and photon production inside jets by the ALEPH experiment~\cite{Buskulic:1995au}. 
These measurements were performed without applying a photon isolation criterion, 
and the ALEPH measurement used a democratic clustering approach~\cite{Glover:1993xc,GehrmannDeRidder:1997gf}  
to combine photons into hadronic jets. Any other measurement 
of final state photon observables at colliders uses a fixed isolation criterion, which has thus no differential sensitivity on  photon 
fragmentation functions. The OPAL measurement was used  to 
discriminate between models for the photon fragmentation function~\cite{Owens:1986mp,Gluck:1992zx} and subsequently  as constraint in their 
fit (BFG parametrization,~\cite{Bourhis:1997yu}). 
In direct comparison~\cite{GehrmannDeRidder:1998ba}, it turns out that the  ALEPH measurement provides higher sensitivity to the
fragmentation function at large longitudinal momentum transfer, which is relevant for the contribution to isolated photon cross sections, and indicate that the 
BFG parametrization slightly underestimates this region. Given the sparse amount of data, it is not possible to quantify a parametric error on the photon 
fragmentation functions. 

For the fixed cone isolation (\fixIso)  procedure, QCD corrections have been computed to NLO for isolated photon and photon-plus-jet production, and implemented in 
the JETPHOX code~\cite{Catani:2002ny}. These predictions use the BFG parametrization~\cite{Bourhis:1997yu} for the photon fragmentation functions. Where 
appropriate, JETPHOX NLO QCD results are displayed in the following for comparison. 

In order to investigate the dependence on the parameter choices for both the dynamical (\dynIso) and the hybrid cone isolation (\hybIso) procedures, 
we use the fiducial cross section definition 
of the $13~\TeV$ ATLAS $\gamma+\mathrm{jet}$ data~\cite{Aaboud:2017kff} (see Section \ref{subsec:GJ_13_A} below).   The photon has to have a transverse momentum $p_T^\gamma > 125~\GeV$ and a rapidity $|y^\gamma| < 2.37$, excluding the barrel--endcap region $[1.37,1.56]$. Each event is required to contain at least one jet, defined through the anti-$k_T$ algorithm~\cite{Cacciari:2008gp} with $R^j=0.4$, with transverse momentum $p_T^j > 100~\GeV$ and rapidity $|y^j| < 2.37$. A jet must have a separation from the photon axis of $R^{\gamma j}>0.8$.

We  compute the theory predictions at NLO, using the NNPDF3.1 PDF set~\cite{Ball:2017nwa}, and both the renormalization and factorization scale are chosen to be equal to the photon transverse momentum. The theoretical uncertainty arising from the scale choice is estimated by means of a seven-point scale variation:
\begin{align}
    \mu_R &= a\,p_T^\gamma\,,   &   
    \mu_F &= b\,p_T^\gamma\,,
\end{align}
where $a,b\in(\frac{1}{2},1,2)$ and we exclude the pairs $(a,b)=(\frac{1}{2},2)$ and $(a,b)=(2,\frac{1}{2})$. For the \fixIso predictions using JETPHOX we superimpose each of the resulting seven scale combinations with a variation of the fragmentation scale $\mu_A$ around a central scale of $p_T^\gamma$
by factors of $\frac{1}{2},1,2$. We observe that the variation of $\mu_A$ has a much smaller impact than the variation of the other two scales. 
\begin{figure}[!t]          
    \centering
    \includegraphics[scale=1.2,angle=270]{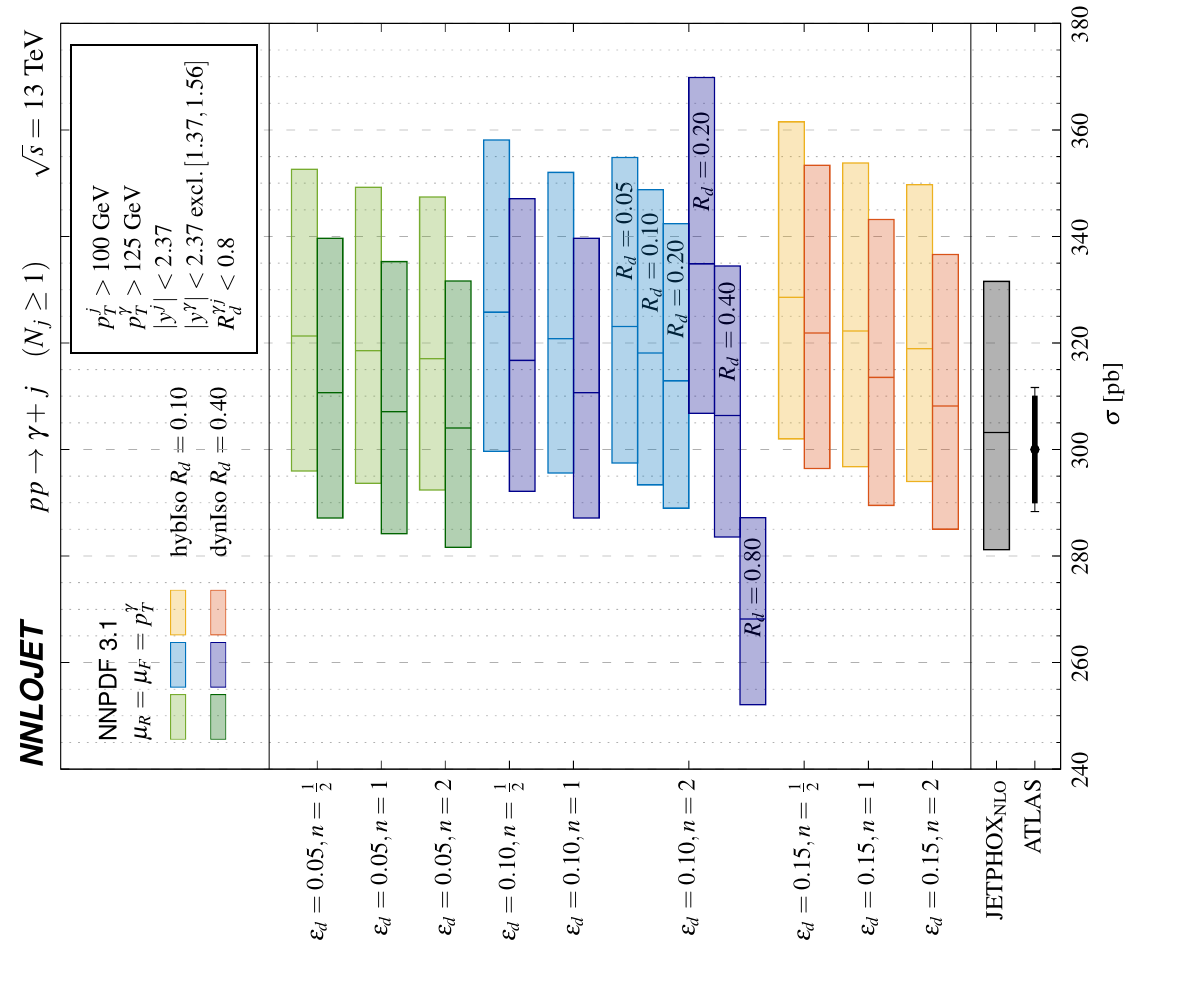}
    \caption{Total cross section for different parameter choices ($\varepsilon_d=0.05,0.1,0.15$, $n=\frac{1}{2},1,2$) for both dynamical photon isolation (\dynIso)~\protect{\cite{Frixione:1998jh}} (dark colours) and the dynamical cone part of the hybrid photon isolation (\hybIso) (light colours). The default cone size for \dynIso is $R_d=0.4$, while for \hybIso it is $R_d=0.1$. For the specific parameter choice  $\varepsilon_d=0.1, n=2$ we also investigate variations of the cone size by factors $\frac{1}{2}$ and $2$. The fixed cone parameters of the hybrid isolation are chosen according to the ATLAS measurement~\protect{\cite{Aaboud:2017kff}}. The 
    \fixIso~prediction (using JETPHOX~\cite{Catani:2002ny}) and the ATLAS measurement are shown for comparison.}%
    \label{fig:ISO_param_dep}
\end{figure}
 
The dynamical cone parameters $\varepsilon_d$ and $n$ are varied in the following ranges:
\begin{align}
    \varepsilon_d &\in (0.05,0.1,0.15)\,,  &   n &\in \left(\tfrac{1}{2},1,2\right)\,.
\end{align}
For these variations, the cone size of the dynamical cone is kept fixed at $R_d = 0.4$ for the standard dynamical isolation and at $R_d = 0.1$ for the hybrid isolation. The dependence on the dynamical
cone size is investigated for  fixed 
$(\varepsilon_d,n) = (0.1,2)$, by taking  $R_d \in (0.2,0.4,0.8)$ for the dynamical isolation and $R_d \in (0.05,0.1,0.2)$ for hybrid isolation. 
In the case of the hybrid isolation the parameters for the outer fixed cone are fixed at
\begin{align}
    R &= 0.4\,, &   
    E_T^\mathrm{thres} &= 10~\GeV\,,    &   
    \varepsilon &= 0.0042\,,
\end{align} 
as in the experimental measurement~\cite{Aaboud:2017kff}. The results are shown in figure~\ref{fig:ISO_param_dep}.  
We observe a reduced dependence on the technical parameters of the dynamical cone when going from dynamical  to hybrid isolation. This reduction is most pronounced for variations of the
cone size $R_d$. This is to be expected, as in the dynamical isolation the dynamical cone defines the actual catchment area for the photon isolation in the calculation, while in the hybrid isolation this is accounted for by the outer fixed cone.
\begin{figure}[!t]
    \centering
    \includegraphics[width=\textwidth]{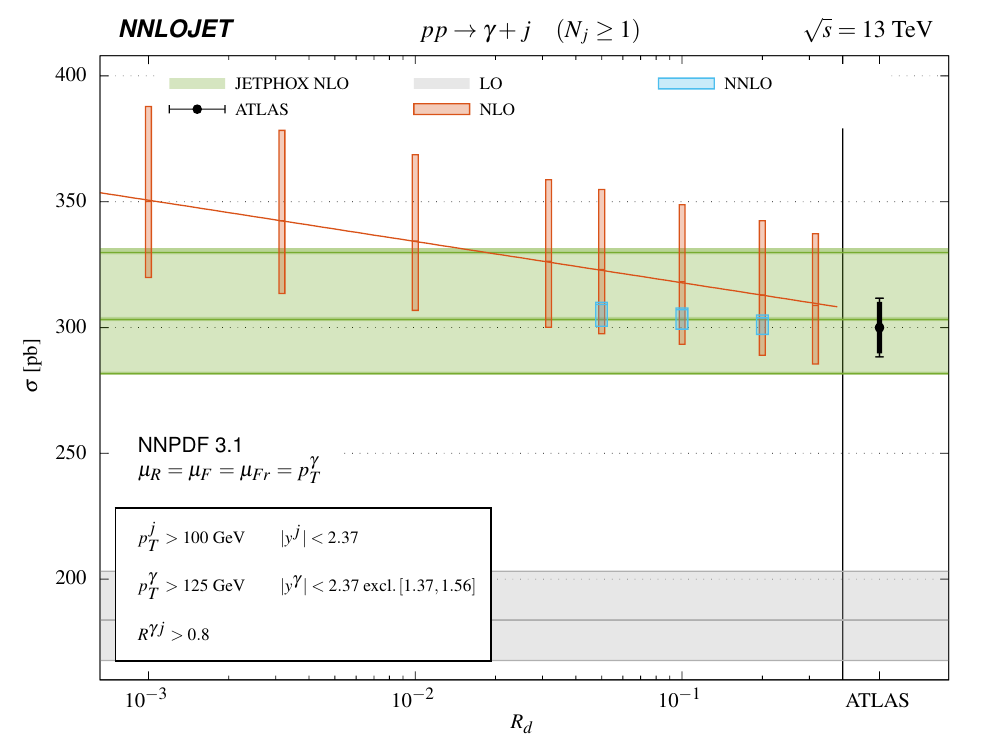}
    \caption{Dependence of the total cross section for photon + jet production (ATLAS 13~TeV measurement~\protect\cite{Aaboud:2017kff}) 
    at NLO and NNLO on the conesize $R_d$ of the inner dynamical cone used in the hybrid isolation procedure. 
    All other isolation parameters are fixed: $\varepsilon_d=0.1$, $n=2$, $R=0.4$, $E_T^\mathrm{thres}=10~\GeV$, $\varepsilon=0.0042$. 
    The line is a fit of a function with form $f(R_d)=a\cdot\log(1/R_d)+b$ to the NLO prediction at the central scale. 
    The NLO prediction for fixed cone isolation is obtained with the BFG parametrization~\protect\cite{Bourhis:1997yu} of the 
    photon fragmentation functions, and computed with the JETPHOX code~\protect\cite{Catani:2002ny}. 
    Its uncertainty band contains only variations of the factorization and renormalization scales, 
    while the small fragmentation scale uncertainty is superimposed on the central and extremal values. The LO prediction is independent on the isolation procedue.}%
    \label{fig:ISO_param_dep_R_d}
\end{figure}
    
Although overlapping within their respective scale uncertainties, the predictions using hybrid isolation display a tendency to fall systematically above the predictions obtained 
using dynamical isolation for identical values of ($\varepsilon_d,n)$, which can be understood from more real radiation events being admitted in the hybrid isolation procedure.
The JETPHOX~\cite{Catani:2002ny} fixed cone isolation prediction is slightly below the bulk of the predictions for both hybrid and dynamical cone isolation, indicating that 
the BFG parametrization~\cite{Bourhis:1997yu} of the photon fragmentation functions amounts to a somewhat smaller amount of photon 
yield in association with partons inside the isolation cone than what is admitted by these prescriptions. 
    
A residual dependence on the cone size $R_d$ in the hybrid isolation remains. For dynamical isolation, as $R_d$ approaches zero, one expects the cross section to diverge as $\sigma\sim\log(1/R_d)$, following from the  factorisation of the cross section in the photon--quark collinear limits at NLO\@. The all-order structure of the leading logarithmic terms is 
understood for a fixed-cone isolation~\cite{Catani:2002ny} and dynamical cone isolation~\cite{Balsiger:2018ezi}. At subleading level, 
non-global logarithms~\cite{Balsiger:2018ezi} appear in either prescription, preventing up to now an understanding of their all-order structure. 

While the outer fixed cone vetoes hard quarks in the vicinity of the photon, relatively soft quarks, that is with $p_T^q<E_T^\mathrm{max}$, are in principle allowed within the fixed cone. The dynamical inner cone prevents them from becoming collinear to the photon. When we shrink the dynamical cone, at some point we will start to probe the quark--photon collinear limit, leading a logarithmic rise,
which at NLO amounts to a single power of $\log(R_d)$. We checked this by extending the scan over the cone size $R_d$ in the hybrid isolation to values as low as $R_d=10^{-3}$. The result is shown in figure~\ref{fig:ISO_param_dep_R_d}. 

The exact limit $R_d\to 0$ corresponds to no photon isolation, resulting in a fully inclusive photon cross section. In this case, the fragmentation contribution to the cross section 
has to be included, which contains a negative and divergent mass-factorisation counter-term from the quark-to-photon fragmentation function~\cite{Laermann:1982jr}, thereby compensating the 
$\log(R_d)$-divergence and yielding a finite result for the cross section. From this cancellation, we can conclude that there exists a finite (but unknown) value of $R_d$ for which the hybrid isolation procedure 
(without fragmentation contribution) should produce  exactly the same results as the fixed-cone isolation (with fragmentation contribution), since the vetoed 
real radiation cross section inside $R_d$ and the negative fragmentation counter-term  exactly compensate each other. The \fixIso result obtained 
with JETPHOX contains this counter-term, together with  the non-perturbative quark-to-photon fragmentation function from which it is inseparable. Comparison of  
the  \hybIso and \fixIso results (and taking into account that the BFG photon fragmentation function parametrization in 
JETPHOX likely underestimates~\cite{GehrmannDeRidder:1998ba} the isolated region) indicates that the exact compensation takes place at around $R_d=0.1$, or even above. 
 
In the following, we use $R_d=0.1$ throughout as default value for hybrid isolation. Smaller values will start probing the quark--photon collinear divergence and 
are disfavoured by the comparison with the  \fixIso results. 
Larger values would violate the condition $R_d^2\ll R^2$, imposed on the relative cone sizes in the hybrid isolation. 
It has to be remembered that the hybrid isolation is an approximation to the fixed-cone isolation used in
the experimental measurements. It reproduces the correct functional dependence on $R$, but induces potentially a small $R$-independent shift on
the cross sections from discarding the collinear fragmentation contributions. The potential magnitude of this shift can be estimated by comparing the 
NLO \hybIso~prediction at  $R_d=0.1$ with the NLO \fixIso~JETPHOX prediction, which is obtained with the BFG parametrization of the photon fragmentation functions and 
predicts a cross section lower by 4.6\%, see Figure~\ref{fig:ISO_param_dep_R_d}.
 This discrepancy likely overestimates the shift, given the effect of the BFG parametrization in the isolated photon region, such that we 
can assume its magnitude to be a conservative upper bound on the residual uncertainty associated with the photon isolation prescription in the theoretical predictions. 
A recent in-depth comparison of the different photon isolation procedures and their uncertainties can be found elsewhere~\cite{Amoroso:2020lgh}. 

At NNLO, the divergent behaviour in the  $R_d\to 0$ limit becomes more involved, containing both $\log^2(R_d)$ and $\log(R_d)$ terms. Resolving the NNLO  
$R_d$-dependence over the full range of values of   Figure~\ref{fig:ISO_param_dep_R_d} is prohibitively expensive in terms of computation time and numerical stability. To illustrate 
the behaviour in the vicinity of the default value  $R_d=0.1$ we display the NNLO cross sections for  $R_d=0.05$ and  $R_d=0.2$, observing that the  $R_d$-dependence
in this region is weaker than at NLO, decreasing from a $(+1.6,-1.7)\%$ variation 
 to a $(+0.9,-1.3)\%$ variation. Following the arguments given above, an exact matching 
of fragmentation counter term and inner hybrid isolation cone should be attained for some value $R_d<R$, and the comparison between 
NLO \fixIso and \hybIso~predictions suggests that this value can not be too small. The variation around the default value  $R_d=0.1$ that is observed at NNLO can thus be
considered a reasonable estimation of the residual uncertainty associated with the hybrid isolation procedure.

\section{Calculation of NNLO corrections}%
\label{sec:calc}

The NNLO QCD corrections to photon production at large transverse momentum receive three types of parton-level contributions: the two-loop 
corrections to photon-plus-one-parton production (double virtual, VV), the virtual corrections to photon-plus-two-parton production (real--virtual, RV), 
and the tree-level photon-plus-three-parton production (double real, RR). The matrix elements are known as closed 
analytic expressions for the  VV~\cite{Anastasiou:2001sv,Bern:2003ck}, RV~\cite{Signer:1995np,Signer:1995a} and RR~\cite{DelDuca:1999pa} processes. 
All three types of contributions contain infrared singularities from the loop integrals, or from soft and collinear real emissions, which cancel only 
once the processes are summed up, and mass factorization is performed on the incoming partons. The numerical implementation  of the NNLO corrections
therefore requires a subtraction method that extracts the infrared-singular configurations from all contributions and combines them to 
yield finite expressions that are suitable for numerical evaluation. We employ the antenna subtraction method at NNLO~\cite{GehrmannDeRidder:2005cm,Daleo:2006xa,Currie:2013vh},
which is implemented in the \nnlojet framework. This parton-level event generator code supplies the computational infrastructure (phase space, event analysis), 
 the building blocks of the subtraction terms 
 (antenna functions~\cite{GehrmannDeRidder:2005hi,GehrmannDeRidder:2005aw,Daleo:2009yj,Gehrmann:2011wi,GehrmannDeRidder:2012ja}), as well as 
 routines for testing and validation. The NNLO  antenna subtraction terms for the photon-plus-jet process are 
 very similar to the ones derived for $Z$+jet production~\cite{Ridder:2015dxa,Ridder:2016nkl}, which 
we used as a template for their construction. Predictions for isolated photon production at large transverse momentum are obtained directly by dropping the jet reconstruction 
requirement. 

The implementation of the RV and RR matrix elements was validated numerically to machine precision
against the OpenLoops code~\cite{Cascioli:2011va,Lindert:2017olm} at the level of phase space points, 
and against SHERPA~\cite{Gleisberg:2008ta,Hoeche:2012yf,Siegert:2016bre}
at the cross section level for LO photon-plus-three-jet and 
NLO photon-plus-two-jet final states within integration errors to sub-per-cent accuracy. 
The MCFM-based calculation of NNLO corrections to direct photon and 
photon-plus-jet production~\cite{Campbell:2016lzl,Campbell:2017dqk} uses a dynamical cone isolation. The detailed 
comparison with these results is described in the following sections in the context of 
the description of the 8 TeV measurements from ATLAS~\cite{Aad:2016xcr} and CMS~\cite{Khachatryan:2015ira}.

For the  numerical predictions throughout this paper, we use the  NNPDF3.1~\cite{Ball:2017nwa} PDF set and apply a hybrid photon isolation (\hybIso) procedure, with outer-cone parameters 
matching the experimental photon isolation criteria. Scale uncertainties are estimated using a seven-point scale variation as in 
section~\ref{subsec:ISO_param_dep}. The electromagnetic coupling is taken in the 
$G_\mu$-scheme as $\alpha_\mathrm{em}^{G_\mu}=1/132.232$. 

\section{Isolated photon production}%
\label{sec:isolated}

Isolated photon cross sections are defined through kinematical selection cuts on the observed photon only. By  requiring
a minimal transverse momentum of the photon, they imply the existence of a partonic recoil. Consequently, 
predictions for isolated photon production are obtained from the photon-plus-jet calculation by simply
dropping the requirement of observing a jet. Experimental measurements of photon production have been performed 
since the early days of hadron colliders~\cite{Diakonou:1979sv,Anassontzis:1982gm,Angelis:1989zv,Albajar:1988im,Alitti:1991yk,Aaltonen:2017swx,Abazov:2005wc}.
Measurements of isolated photon production at 
ATLAS~\cite{Aad:2011tw,Aad:2016xcr,Aaboud:2017cbm}  and CMS~\cite{Chatrchyan:2011ue,Sirunyan:2018gro} are now reaching per-cent level accuracy 
over a large kinematical range. To interpret these precision data 
demands an equally high accuracy on the theory predictions. In the following, we confront the 8 TeV ATLAS data~\cite{Aad:2016xcr} and the 13~TeV 
ATLAS~\cite{Aaboud:2017cbm} and CMS~\cite{Sirunyan:2018gro} data with our newly derived NNLO QCD predictions. By default, we use the hybrid isolation procedure 
described in Section~\ref{sec:photondef}. In order to compare our results with the MCFM calculation of the NNLO corrections, we also replicate the setup of~\cite{Campbell:2016lzl} 
by employing a dynamical cone isolation with the same parameters as used there, confronted to the ATLAS 8~TeV measurements.

\subsection{Comparison with ATLAS 8 TeV measurements and MCFM calculation}%
\label{subsec:iG_8_A}

The ATLAS 8 TeV measurement~\cite{Aad:2016xcr} of isolated photon production is performed in four different regions in rapidity
\begin{equation}
    |y^\gamma| < 0.6, \quad 0.6<|y^\gamma| < 1.37,\quad 1.56<|y^\gamma| < 1.81,\quad 1.81<|y^\gamma| < 2.37\,,
    \label{eq:ATLASrap}
\end{equation}
and  differentially in transverse momentum, with 
a lower cut off $p_T^\gamma > 25~\GeV$. No further cuts are applied.
\begin{figure}[!t]
    \centering
    \includegraphics[scale=1.2]{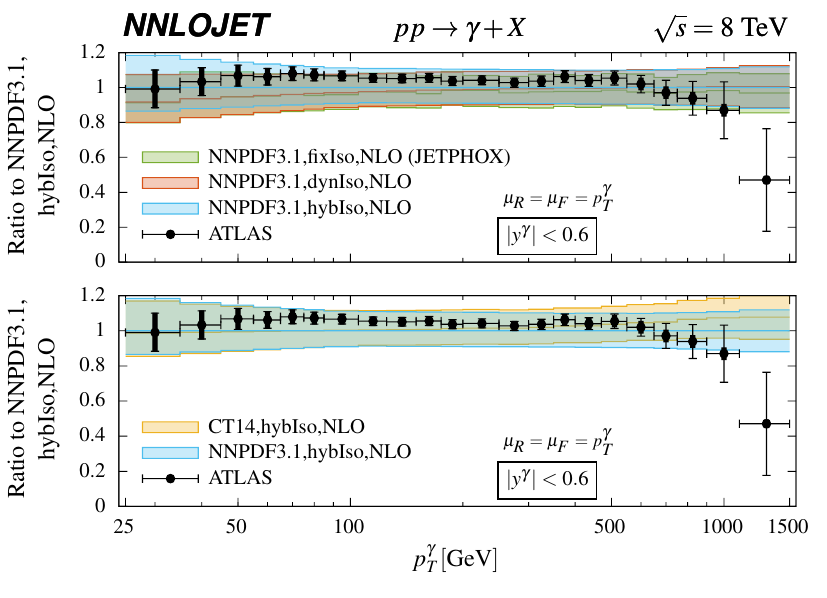}
    \caption{Impact of the choice of the photon isolation criterion (dynamical isolation~\protect{\cite{Frixione:1998jh}} and fixed-cone isolation (produced with 
    JETPHOX~\protect\cite{Catani:2002ny}) versus hybrid isolation, top frame) and PDF set (CT14~\protect{\cite{Dulat:2015mca}} versus NNPDF3.1~\protect{\cite{Ball:2017nwa}}, bottom frame) at NLO, shown as ratio to our default choice of  NNPDF3.1 with hybrid isolation.}
    \label{fig:iG_8_A_etG_1_comp}
\end{figure}        
        
For the theoretical \nnlojet predictions, we set the central renormalization and factorization scale to be equal to the photon transverse momentum $p_T^\gamma$. 
As default, we use the NNPDF3.1~\cite{Ball:2017nwa} PDF set and apply a hybrid photon isolation (\hybIso)
with parameters
\begin{align}
    R_d &= 0.1\,,   &   \varepsilon_d &= 0.1\,, &   n &= 2\,,\nonumber \\   
    R &= 0.4\,, &   E_T^\mathrm{thres} &= 4.8~\GeV\,,   &   \varepsilon &= 0.0042\,,
    \label{eq:ATLASiso}
\end{align}
such that the outer fixed-cone parameters ($R$, $E_T^\mathrm{thres}$, $\varepsilon$) reproduce the photon isolation definition used in the experimental measurement~\cite{Aad:2016xcr}. 

In order to compare with the MCFM calculation~\cite{Campbell:2016lzl}, we replicate the setup used there, with the CT14~\cite{Dulat:2015mca} PDF set and a dynamical cone isolation (\dynIso)
\begin{align}
    R_d &= 0.4\,,   &   \varepsilon_d &= 0.1\,, &   n &= 2\,.
\end{align}         
To investigate the impact of these different settings, we compare the combinations of PDF and isolation procedure at NLO. 
Finally, at NLO we also use JETPHOX~\cite{Catani:2002ny} to compute predictions for fixed cone isolation (\fixIso) with 
the cone parameters of the experimental measurement, and using   the BFG parametrization~\cite{Bourhis:1997yu} of the photon fragmentation functions.

In the upper panel of figure~\ref{fig:iG_8_A_etG_1_comp} we compare the \dynIso and \fixIso predictions to the default setting of \hybIso. All three predictions are obtained using 
 NNPDF3.1 parton distributions.
 We find that the largest differences due to the choice of the isolation procedure occur in the low $p_T^\gamma$ region below approximately $100~\GeV$, while for $p_T^\gamma > 200~\GeV$ the difference is negligible. The cross section obtained with \dynIso or \fixIso (which yield very similar predictions)
is consistently lower than the \hybIso result, as already observed for the total cross section in Figure~\ref{fig:ISO_param_dep}. 
In the lowest bin the deviation in the central value lies just below $10\%$. This discrepancy is in principle consistent at NLO within the scale uncertainty.
 It is noted, however, that unlike scale setting 
effects, the impact of the photon isolation procedure is not compensated at higher orders, such that the difference reflects a genuine systematic shift in the predictions.
 
The lower panel of figure~\ref{fig:iG_8_A_etG_1_comp} compares the \hybIso predictions for NNPDF3.1 and CT14 parton distributions. Here, 
 we observe the opposite kinematical pattern. 
While there is no significant difference at low $p_T^\gamma$, using CT14 leads to a consistently larger cross section compared to NNPDF3.1 for $p_T^\gamma > 200~\GeV$, 
up to almost $8\%$ in the highest bin. This pattern can be traced back to differences in the large-$x$ gluon and antiquark distributions, which produces similar effects 
also in gauge-boson-plus-jet observables~\cite{Gehrmann-DeRidder:2019avi}. 

The NNLO prediction for the ATLAS 8~TeV isolated photon production is computed for our default setting of NNPDF3.1 and hybrid isolation and, in order to numerically compare with the MCFM study~\cite{Campbell:2016lzl}, also for their choice, CT14 and dynamical isolation. 

\begin{figure}[t]
    \centering
    \includegraphics[scale=1.2]{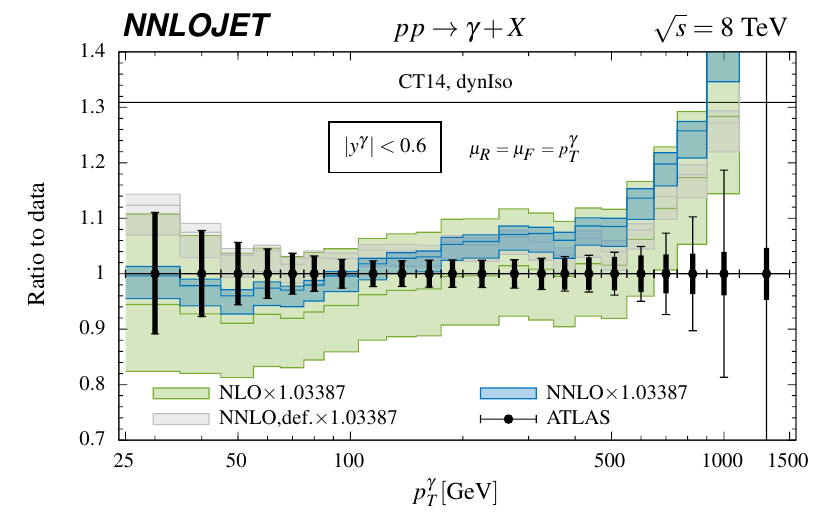}
    \caption{Ratio to ATLAS data~\protect{\cite{Aad:2016xcr}} for the transverse energy/momentum of the photon at NLO and NNLO, using the PDF and isolation procedure choice of MCFM~\protect{\cite{Campbell:2016lzl}}.  The results are rescaled by a factor of $1.03387$ to account for the different choice of $\alpha_\mathrm{em}$.  The NNLO result obtained using the default PDF and isolation procedure choice is shown in grey.}%
    \label{fig:iG_8_A_etG_val}
\end{figure}

We also need to take into account the different  value of  $\alpha_\mathrm{em}(M_Z)=1/127.9$ used there 
for the electromagnetic coupling, while our predictions are obtained in the  $G_\mu$-scheme with $\alpha_\mathrm{em}^{G_\mu}=1/132.232$. 
Since we are only considering the QCD corrections to one-photon amplitudes, the results are directly proportional to $\alpha_\mathrm{em}$ and thus the difference in $\alpha_\mathrm{em}$ can readily be accounted for by a constant rescaling factor 
\begin{align}
    \frac{\alpha_\mathrm{em}(M_Z)}{\alpha_\mathrm{em}^{G_\mu}} &= \frac{1/127.9}{1/132.232} \approx 1.03387\,.
\end{align}
Figure~\ref{fig:iG_8_A_etG_val} shows the ratio to ATLAS data at NLO and NNLO. It corresponds to the lower panel of figure~4 in the MCFM study~\cite{Campbell:2016lzl}, where however the bins below $p_T^\gamma=65~\GeV$ are not displayed.
Compared to the default setting,  we observe a decrease in the low $p_T$ region caused by the dynamical isolation as well as an increase in the high $p_T$ region, due to the use of CT14. By construction, this agrees with our findings from figure~\ref{fig:iG_8_A_etG_1_comp}.  
        
While our NLO results are in full agreement with the MCFM study~\cite{Campbell:2016lzl}, we observe discrepancies at NNLO\@. Especially at low $p_T^\gamma$, our predictions are above the ones obtained 
in~\cite{Campbell:2016lzl}. Moreover, for all values of $p_T^\gamma$ we compute a  scale uncertainty that is slightly larger than the one stated in~\cite{Campbell:2016lzl}. A most recent re-evaluation of the MCFM results~\cite{Campbell} leads to modifications that bring MCFM and our results into mutual agreement within their respective Monte Carlo uncertainties. 
It should be emphasised that the two calculations  rely on independent implementations of the underlying NNLO matrix elements and use completely different methods for the extraction and 
cancellation of infrared singularities among the different subprocesses. Consequently, the observed agreement 
 amounts to a highly non-trivial check for our result as well as for MCFM~\cite{Campbell}. 
\begin{figure}[!t]
    \centering
    \includegraphics[scale=0.9]{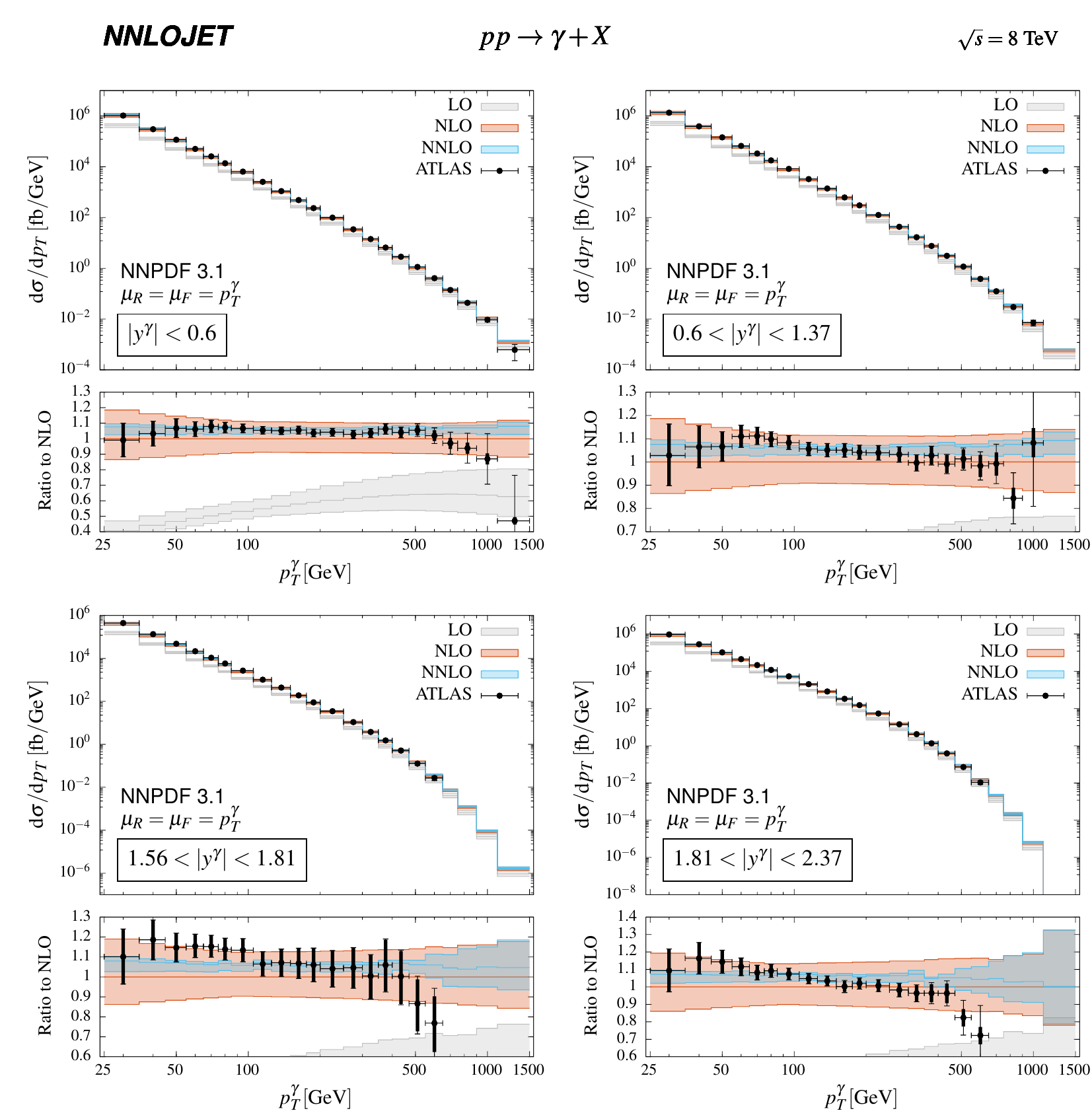}
    \caption{Transverse energy/momentum distribution of isolated  photons at LO, NLO, NNLO in four different rapidity bins, from central (top left) to most forward (bottom right). The results are compared to 8 TeV  ATLAS data~\protect{\cite{Aad:2016xcr}}.
    } 
    \label{fig:iG_8_A_etG}
\end{figure}        
        
Figure~\ref{fig:iG_8_A_etG} shows a detailed comparison of the NNLO predictions obtained with our default setup with the ATLAS 8~TeV data~\cite{Aad:2016xcr}. Compared to NLO, the inclusion of NNLO corrections leads to a substantial reduction of the scale uncertainty on the predictions to less than $(+2,-4)$\% in the bulk of the $p_T^\gamma$ distributions, with slightly larger uncertainty towards the limits of large and small transverse momentum, with the exception of the two forward bins, for which the scale uncertainty grows drastically in the highest $p_T^\gamma$-bins, in which the cross-section drops over several orders of magnitude. Throughout the kinematical range, the NNLO scale uncertainty is at most as large as (and mostly smaller than) the measurement errors. 
The ATLAS data are well-described in normalization and shape for all rapidity ranges. Small deviations observed at the largest transverse momenta are not yet significant within error ranges,
but might indicate the onset of electroweak Sudakov logarithms~\cite{Becher:2013zua,Becher:2015yea}.

\subsection{Comparison with ATLAS 13 TeV measurements}%
\label{subsec:iG_13_A}

The ATLAS 13~TeV isolated photon measurement~\cite{Aaboud:2017cbm} is performed for the same rapidity bins~\eqref{eq:ATLASrap} as used at 8 TeV~\cite{Aad:2016xcr}
with a fixed-cone based isolation and transverse momentum $p_T^\gamma > 125~\GeV$. Compared to the 8~TeV measurement, this larger transverse momentum cut implies 
a reduced sensitivity on the photon isolation prescription. 
\begin{figure}[!t]
    \centering
    \includegraphics[scale=0.9]{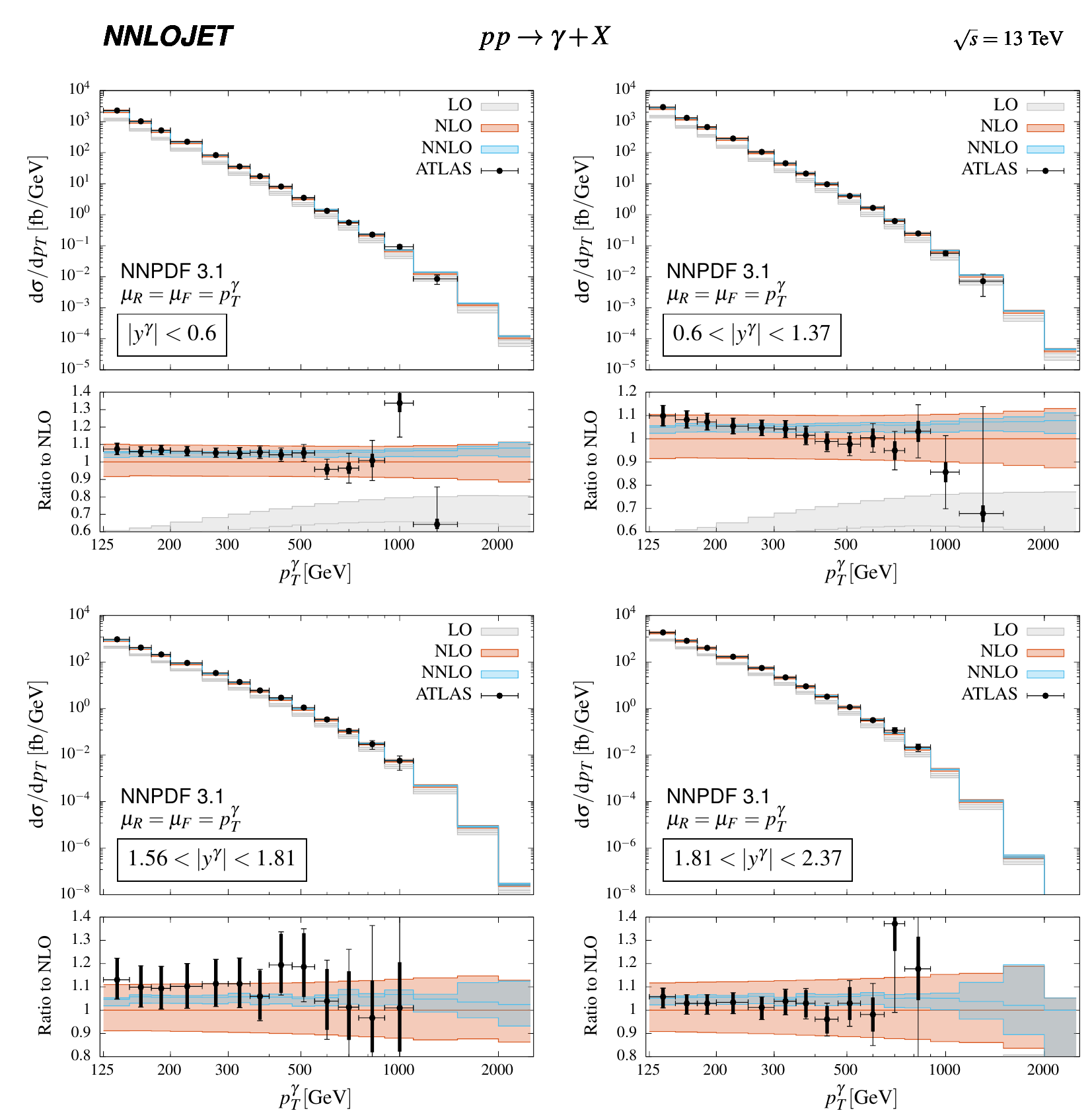}
    \caption{Transverse energy/momentum distribution of isolated photons at LO, NLO and NNLO in four different rapidity bins, from central (top left) to most forward (bottom right). The results are compared to 13 TeV ATLAS data~\protect{\cite{Aaboud:2017cbm}}.}%
    \label{fig:iG_13_A_etG}
\end{figure}
        
For the theoretical predictions, we use NNPDF3.1, and the hybrid isolation procedure with
\begin{align}
    R_d &= 0.1\,,   &   \varepsilon_d &= 0.1\,, &   n &= 2\,,\nonumber \\
    R &= 0.4\,, &   E_T^\mathrm{thres} &= 4.8~\GeV\,,   &   \varepsilon &= 0.0042\,,
    \label{eq:ATLASiso13}
\end{align}
where the parameters for the outer cone correspond to the settings used in the ATLAS measurement. 
Central renormalization and factorization scales are again set equal to the photon transverse momentum $p_T^\gamma$.
Figure~\ref{fig:iG_13_A_etG} shows the four rapidity bins up to $|y^\gamma| = 2.37$, excluding the region $[1.37,1.56]$ and compared to ATLAS data~\cite{Aaboud:2017cbm}.
The NNLO corrections are positive and largely constant over the whole rapidity and $p_T^{\gamma}$ range, increasing the prediction for the central scale by approximately $(5-6)\%$. The scale uncertainty at NLO is around $\pm10\%$ for the central rapidity bin and increases to more than $\pm15\%$ for the more forward bins. At NNLO this uncertainty is significantly reduced to no more than $(+3.2,-5.1)\%$ in all bins, in most bins to even smaller values, except at very large  $p_T^\gamma$ for the last three bins in the the two most forward regions. Here the cross section drops quickly and the scale uncertainty increases. 
    
Overall we observe a very good agreement with the data in most bins. Larger discrepancies are observed only for the highest values of  $p_T^\gamma$, where data and theory remain nevertheless 
consistent within increasing experimental errors. 
In the second rapidity bin,  we observe that the slope of the measured $p_T^\gamma$ distribution is less well described than in the other bins, with the theory prediction being slightly harder than the 
measurement.

\subsection{Comparison with CMS 13 TeV measurements}%
\label{subsec:iG_13_C}

The CMS 13 TeV measurement of isolated photon production~\cite{Sirunyan:2018gro} is performed in four bins in rapidity
\begin{equation}
|y^\gamma| < 0.8, \quad 0.8<|y^\gamma| < 1.44,\quad 1.57<|y^\gamma| < 2.1,\quad 2.1<|y^\gamma| < 2.5\,,
\label{eq:CMSrap}
\end{equation}
and yields photon transverse momentum distributions for $p_T^\gamma>190$~GeV. It uses a fixed-cone isolation procedure. 
\begin{figure}[!t]
    \centering
    \includegraphics[scale=0.9]{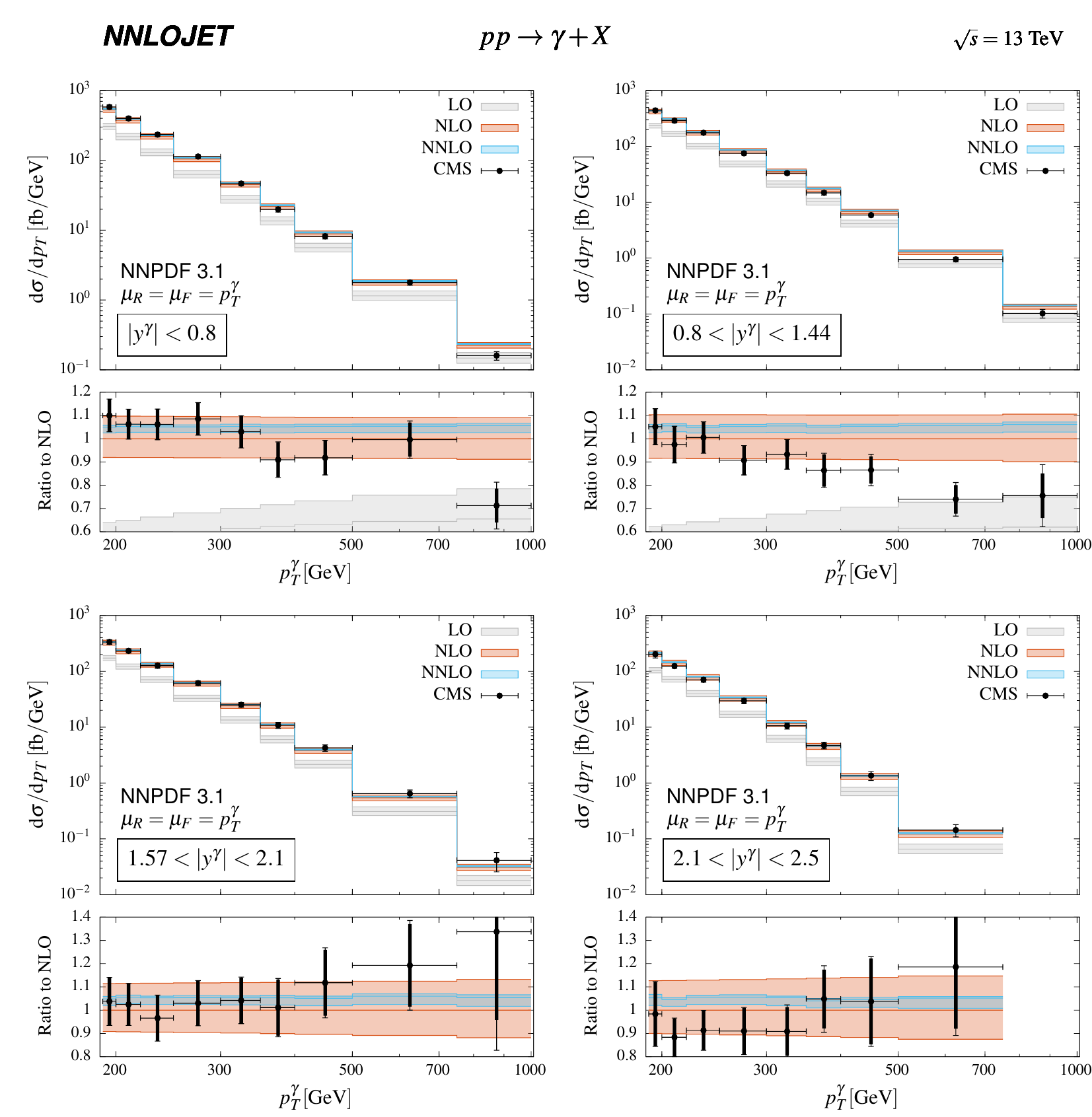}
    \caption{Transverse energy/momentum distribution of the photon at LO, NLO and NNLO in four different rapidity bins, from central (top left) to most forward (bottom right). The results are compared to 13 TeV CMS data~\protect{\cite{Sirunyan:2018gro}}. Note that for the sake of comparison the data has been multiplied by the corresponding rapidity bin-width, as CMS presents the data in double-differential form in $p_T^\gamma,\,y^\gamma$.}
    \label{fig:iG_13_C_etG}
\end{figure}

We compute the theory predictions using NNPDF3.1 with a central scale of $p_T^\gamma$, and use the hybrid isolation parameters      
\begin{align}
    R_d &= 0.1\,,   &   \varepsilon_d &= 0.1\,, &   n &= 2\,,\nonumber \\
    R &= 0.4\,, &   E_T^\mathrm{thres} &= 5~\GeV\,, &   \varepsilon &= 0\,,
    \label{eq:CMSiso}
\end{align}
with the large cone parameters coinciding with the fixed-cone settings used by CMS.         
Figure~\ref{fig:iG_13_C_etG} shows the result in the four rapidity bins up to $|y^\gamma| = 2.5$, excluding the region $[1.44,1.57]$.
Again we find the the NNLO corrections to be positive and largely constant, increasing the NLO predictions by roughly $(4-6)\%$ for the central scale. 
The scale uncertainties are similar as observed in the previous subsection: at NLO approximately $\pm10\%$ for central 
rapidities and growing to $\pm15\%$ for the most forward bin, and no more than $(+1.4,-4.2)\%$ at NNLO\@. 

Most data points agree with the calculation within the respective experimental and theoretical uncertainty, with discrepancies mainly observed in the bins with the largest $p_T^\gamma$. 
Again, the theory prediction for the slope of the $p_T^\gamma$ distribution in the second rapidity bin is harder than what is observed in the experimental data. This effect is even more 
pronounced for the CMS data than for the ATLAS data. Given that ATLAS and CMS display a similar pattern in this region using 13 TeV data, this may point towards the need to reconsider the parton
distributions in kinematical ranges relevant to this distribution.

\subsection{Dependence on photon isolation parameters}%
\label{subsec:iG_13_A_R_dep}

\begin{figure}[!t]
    \centering
    \includegraphics[scale=0.9]{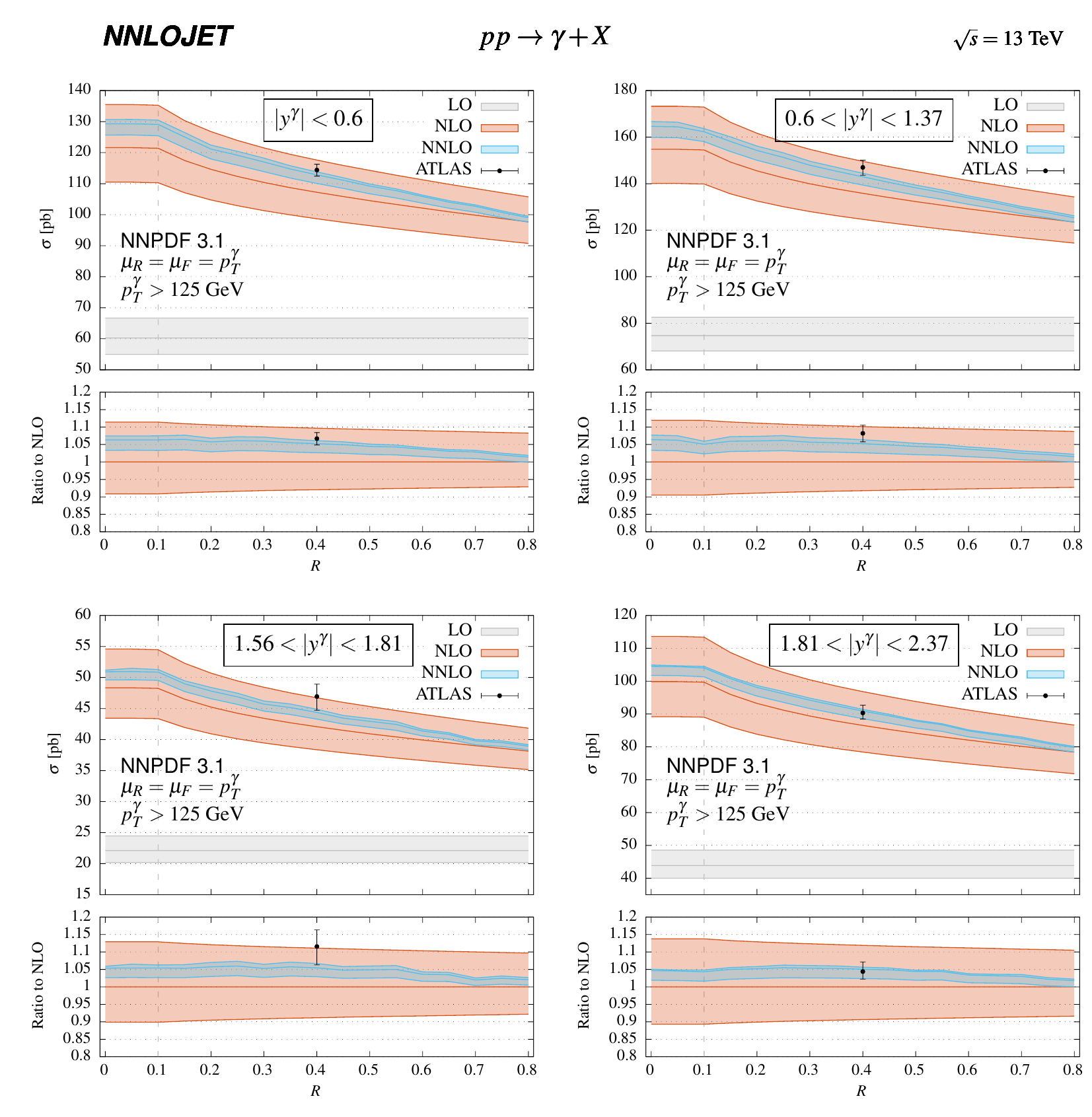}
    \caption{Dependence of the total cross section for inclusive photon production at $13~\TeV$ at LO, NLO and NNLO on the cone size $R$ of the outer fixed cone used in the hybrid isolation procedure, for different regions of the photon rapidity. All other isolation parameters are fixed: $R_d=0.1$, $\varepsilon_d=0.1$, $n=2$, $E_T^\mathrm{thres}=4.8~\GeV$, $\varepsilon=0.0042$. The dashed line marks the cone size $R_d$ of the dynamical cone.  The ATLAS measurement~\protect{\cite{Aaboud:2017cbm}} is performed only for a fixed cone with size $R=0.4$.}%
    \label{fig:iG_13_A_R_dep}
\end{figure}
The hybrid isolation procedure approximates the fixed-cone isolation that is used in the experiment through a theory prescription that vetos collinear quark--photon configurations 
and eliminates the contribution from the photon fragmentation functions. This behaviour is obtained by applying a dynamical isolation procedure inside a small inner cone of radius $R_d$, concentric to 
the larger isolation cone of radius $R$. As a consequence, some amount of hadronic energy inside the inner cone is not properly treated in the theory calculation, resulting in a 
systematic mismatch on the cross section prediction. The resulting small offset will vary with the $E_T^\mathrm{max}$ that is used in the experimental isolation, but is independent on $R$, as long as $R>R_d$, since 
it affects only parton radiation inside $R_d$. Consequently, our calculation in hybrid isolation can predict the variation of the isolated photon cross section under changes of the size 
of the isolation cone $R>R_d$. The neglected photon fragmentation process contributes only inside $R_d$, and potentially leads to an $R$-independent offset on the normalization of the 
cross section, as discussed in detail in Section~\ref{subsec:ISO_param_dep} above.
        
As a test case for the $R$-dependence, we consider the  isolated photon cross section of the ATLAS 13~TeV measurement~\cite{Aaboud:2017cbm} (discussed in section~\ref{subsec:iG_13_A}
above) integrated in $p_T^\gamma$, for the four different rapidity bins~\eqref{eq:ATLASrap}. We use our default setup with hybrid isolation parameters as in~\eqref{eq:ATLASiso13}, varying only $R$. 
Figure~\ref{fig:iG_13_A_R_dep} displays the $R$-dependence of the cross sections at different perturbative orders. For LO, the cross section is constant, since the number of partons is 
insufficient to trigger the cone-based isolation. 
We vary the fixed cone size between $R=0$ to $R=0.8$, and observe that the $R$-dependence is very similar in all four rapidity bins. 
As expected, we see a decrease of the cross section when going to higher values of $R$, as an increasing portion of the phase space for the extra QCD radiation is vetoed. This decrease is slightly stronger 
at NNLO than at NLO, likely due to the improved description of extra radiation with increasing number of external partons. The scale uncertainty on the NNLO cross section is not larger than $(+1.3,-2.9)\%$. Once the cone size of the outer fixed cone becomes smaller than the size of the dynamical cone  $R<R_d$, the hybrid isolation prescription becomes 
largely identical to a dynamical cone isolation with cone size $R_d$, since the catchment area of the fixed cone falls fully inside the dynamical cone. 
 This can be seen in the figures for $R<0.1$, with a near-flat cross section 
indicating that the behaviour is essentially dictated by the dynamical cone isolation step.

\section{Photon-plus-jet production}%
\label{sec:jet}

The measurement of hadronic jets produced in association with an isolated photon allows for the 
 direct reconstruction of the leading-order kinematics of the underlying two-to-two scattering process, thereby constraining in particular the momentum fractions of the incoming 
 partons. Following earlier studies at the Tevatron~\cite{Abazov:2008er,D0:2013lra}, ATLAS~\cite{Aad:2013gaa,Aaboud:2016sdm,Aaboud:2017kff} 
 and CMS~\cite{Chatrchyan:2013oda,Chatrchyan:2013mwa,Khachatryan:2015ira,Sirunyan:2018gro} provided precision measurements of 
 photon-plus-jet production over a large kinematical range. 
 
 The interpretation of these data, and their potential usage in extraction of parton distribution functions, requires precise theory predictions.
 Our NNLO corrections for photon-plus-jet production 
 are compared to the 8 TeV CMS data~\cite{Khachatryan:2015ira} and to the 13 TeV ATLAS~\cite{Aaboud:2017kff} and CMS~\cite{Sirunyan:2018gro} data. 
 By default, the hybrid isolation procedure is applied. For comparison with the MCFM calculation, we also replicate the setup of~\cite{Campbell:2017dqk} 
 using a dynamical cone isolation with the same parameters as chosen there, confronted with the CMS 8 TeV measurements. 
 
\subsection{Comparison with CMS 8 TeV measurements and MCFM calculation}%
\label{subsec:GJ_8_C}

\begin{figure}[!t]
    \centering
    \includegraphics[scale=1.2]{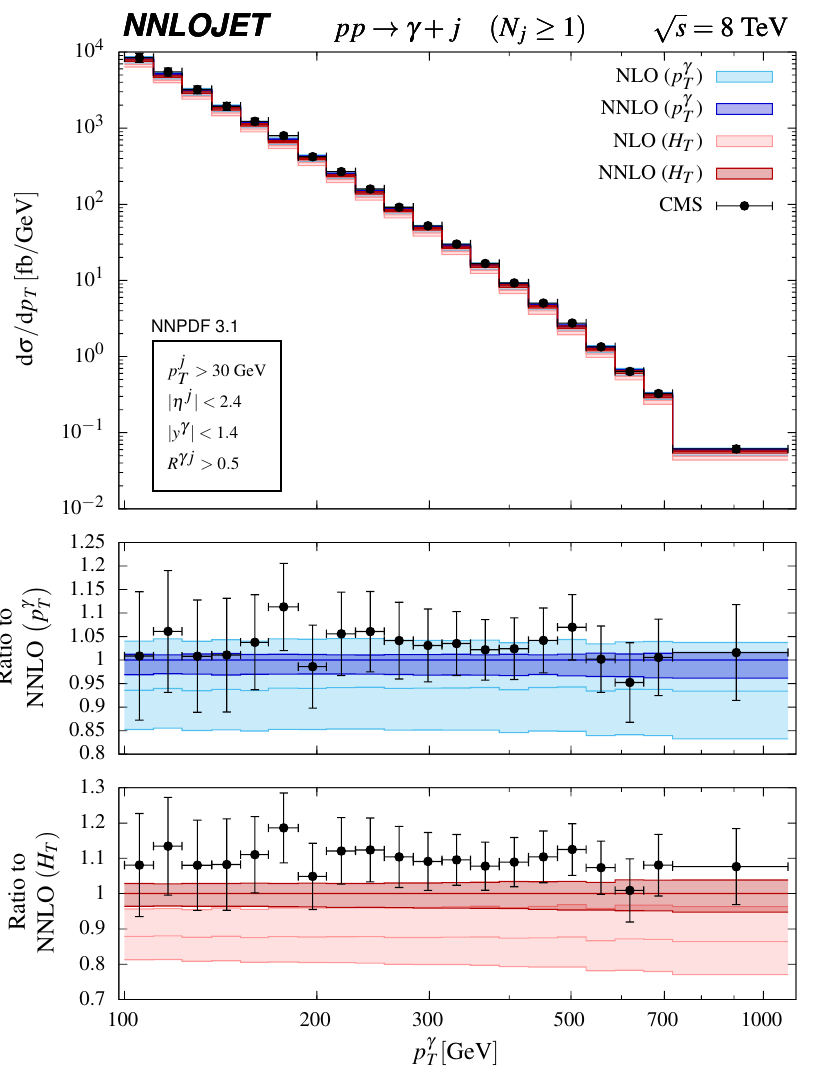}
    \caption{Transverse energy/momentum distribution of the photon at NLO and NNLO for different central scale choices: $p_T^\gamma$ in blue, $H_T$ in red. The calculations are carried out using the NNPDF3.1 NNLO PDF set~\protect{\cite{Ball:2017nwa}} and the hybrid isolation procedure. The results are compared to 8~TeV CMS data~\protect{\cite{Khachatryan:2015ira}}.}%
    \label{fig:GJ_8_C_etG}
\end{figure}        
The CMS measurement of photon-plus-jet production uses the anti-$k_T$ algorithm~\cite{Cacciari:2008gp} with a radius parameter of $R^j=0.5$ to perform the jet clustering. The following cuts in transverse momentum and pseudorapidity 
are applied to the jets:
\begin{equation}
p_T^j > 30~\GeV,\qquad |y^j|<2.4.
\label{eq:CMSjet}
\end{equation}
The measurement is inclusive on the jet multiplicity, meaning that events are retained if they contain at least one jet passing these cuts. 
Photons are identified with a fixed-cone isolation and 
must be separated in azimuth and pseudorapidity from the jet axis by  $R^{\gamma j} > 0.5$. 
Their transverse momentum distribution for $p_T^\gamma > 100~\GeV$ is measured in the central rapidity region $|y^\gamma| < 1.4$.

Our calculation is performed with the default setting, using NNPDF3.1 parton distributions and hybrid isolation with the same parameters as in~\eqref{eq:CMSiso}, matching the fixed-cone settings 
of the CMS measurement~\cite{Khachatryan:2015ira}. We use two different values for the central scale: $\mu_R=\mu_F=p_T^\gamma$ and 
$\mu_R=\mu_F=H_T$, where $H_T$ is defined as the scalar sum of the transverse momenta of all final state partons and the photon. A central scale at $p_T^\gamma$ is our standard choice, while $H_T$ has been used 
in the MCFM calculation~\cite{Campbell:2017dqk}. The results are shown in figure~\ref{fig:GJ_8_C_etG}. For both central scale choices we find the NNLO corrections to be positive. For 
$H_T$, they are typically  $(14-16)\%$, which is considerably larger than the $(6.0-7.1)\%$ corrections obtained for  $p_T^\gamma$. 
While in both cases the scale uncertainty at NLO is up to $\pm11\%$, it is decreased at NNLO to up to $(+1.7,-3.8)\%$ for $p_T^\gamma$ and up to $(+3.9,-5.3)\%$ for $H_T$. In terms of 
perturbative stability,  $p_T^\gamma$ appears thus to be sightly favourable as the central scale choice. 
Although within experimental and theoretical uncertainties both scale choices are consistent with the data, the calculation carried out using $p_T^\gamma$ yields a better description of data, while 
the predictions using $H_T$ are below the data in almost all bins.

\begin{figure}[!t]
    \centering
    \includegraphics[scale=1.2]{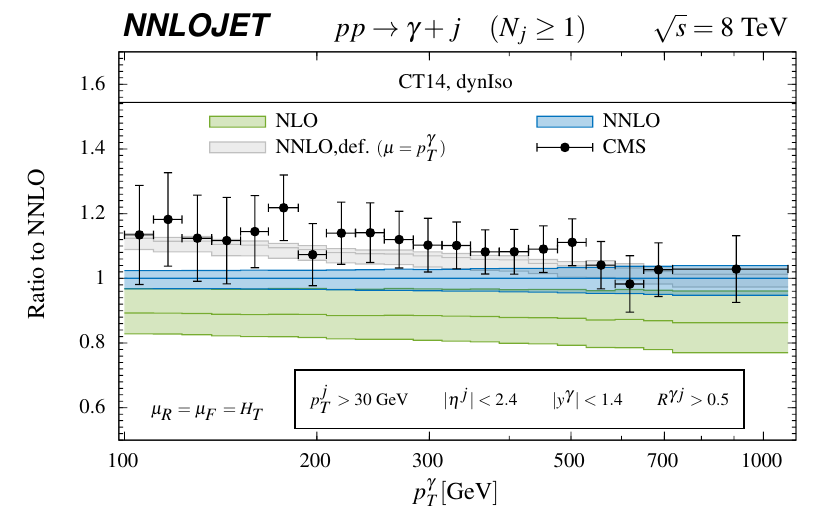}
    \caption{Ratio to NNLO for the transverse energy/momentum of the photon at NLO and NNLO, using the PDF and isolation procedure choice of MCFM~\protect{\cite{Campbell:2017dqk}}. The default NNLO result is shown in grey.}%
    \label{fig:GJ_8_C_etG_val}
\end{figure}
To compare our result  to the MCFM calculation of the NNLO corrections~\cite{Campbell:2017dqk}, we replicate the setup used there, with  CT14 PDF, dynamical cone isolation
with
\begin{align}
    R_d &= 0.4  &   \varepsilon_d &= 0.025\,,   &   n &= 2\,,
\end{align}     
and $H_T$ as the central scale choice. The NLO and NNLO results are shown in figure~\ref{fig:GJ_8_C_etG_val}, which reproduces the lower panel of figure~2 in the MCFM study. 
We observe a good agreement with the result presented in~\cite{Campbell:2017dqk}, which provides an important cross-check on both calculations, which were performed with 
completely different methods, and which rely on fully independent implementations. The specific aspects of the MCFM calculation that required a re-evaluation of the isolated photon results, discussed in section~\ref{subsec:iG_8_A} above, are not expected to have a significant impact on the results for the photon+jet process~\cite{Campbell} that are compared here. 

\subsection{Comparison with ATLAS 13 TeV measurements}%
\label{subsec:GJ_13_A}

Detailed measurements of kinematical distributions in photon-plus-jet production were performed by the ATLAS collaboration~\cite{Aaboud:2017kff}, based on data 
taken at 13~TeV. The study uses the anti-$k_T$ algorithm with $R^j=0.4$ to identify the jets, and the following parameters for the fixed-cone based photon isolation:
\begin{align}
    R &= 0.4\,, &   E_T^\mathrm{thres} &= 10~\GeV\,,    &   \varepsilon &= 0.0042\,,  \label{eq:ATLAS13isoA}
\end{align}         
which only differ in the threshold energy $E_T^\mathrm{thres}$ from the ones used in the inclusive photon measurement discussed in Section~\ref{subsec:iG_13_A}. We take the same parameters for the dynamical cone of the hybrid isolation procedure as above:
\begin{align}
   R_d &= 0.1\,,   &   \varepsilon_d &= 0.1\,, &   n &= 2\,. \label{eq:ATLAS13isoB}
\end{align}
 The fiducial event selection cuts
are as follows: 
\begin{eqnarray}
    &p_T^j > 100~\GeV,\quad |y^j|<2.37, \nonumber \\  
    &p_T^\gamma > 125~\GeV,\quad (|y^\gamma|<1.37~ \text{or}~ 1.56 < |y^\gamma|<2.37), \quad
     R^{\gamma j}>0.8.
    \label{eq:ATLAScuts13}
\end{eqnarray}
The measurement requires that at least one jet passes the above jet cuts, and is thus inclusive on the number of jets. Distributions involving the jet kinematics always refer to the leading (in transverse momentum) jet. For some observables examining the photon--jet system, additional cuts are imposed:
\begin{align}
    |y^\gamma + y^j| &< 2.37\,, &   m^{\gamma j} &> 450~\GeV\,, &   |\cos\theta^*| &< 0.83\, ,
    \label{eq:ATLAShighmass}
\end{align}
where
\begin{align}\label{eq:costhetastar}
    \cos\theta^* &= \tanh\frac{\Delta y^{\gamma j}}{2}\,,
\end{align}     
with $\Delta y^{\gamma j}$ being the rapidity difference between the photon and the leading jet. In the centre-of-momentum system of the underlying two-to-two Born process $\theta^*$ corresponds to the scattering angle. In the following, the fiducial selection cuts are explicitly indicated in the figures.                        

\begin{figure}[!t]
    \centering
    \includegraphics[scale=1.2]{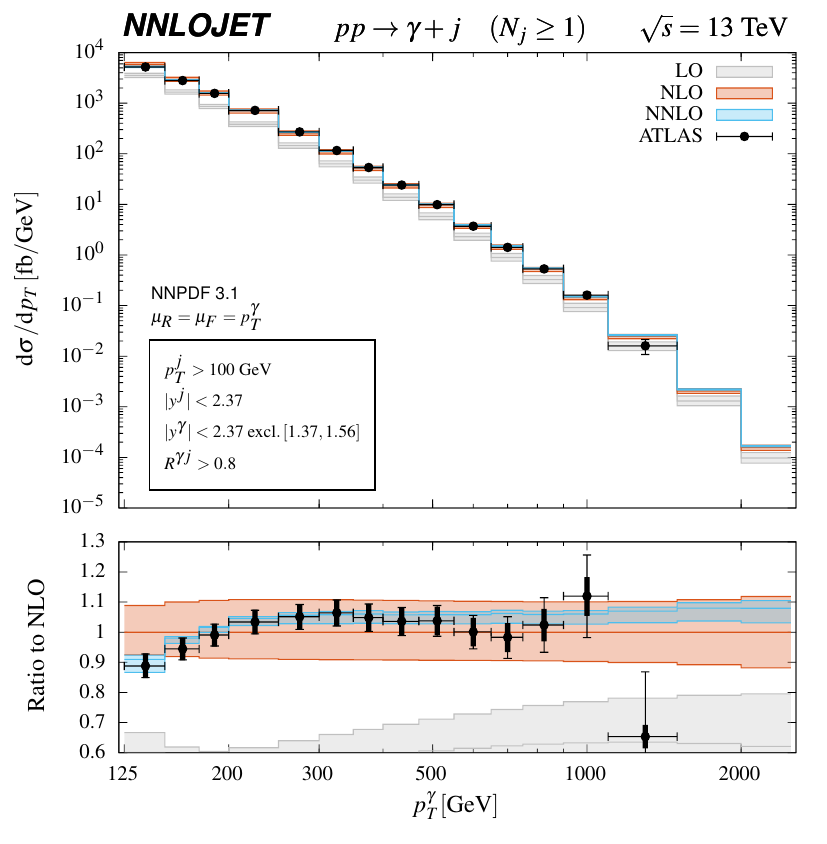}
    \caption{Transverse momentum distribution of the photon in photon-plus-jet events, at LO, NLO and NNLO\@. The predictions are compared to ATLAS data~\protect{\cite{Aaboud:2017kff}}.}%
    \label{fig:GJ_13_A_etG}
\end{figure}

We compute the theory predictions in our default setting, with NNPDF3.1 and hybrid isolation, using the parameters listed in~\eqref{eq:ATLAS13isoA} and \eqref{eq:ATLAS13isoB} and with $p_T^\gamma$ as 
central scale. 
The transverse momentum distribution of the photon  $p_T^\gamma$ is  
compared to the ATLAS data~\cite{Aaboud:2017kff} in figure~\ref{fig:GJ_13_A_etG}. 
We see that going from NLO to NNLO leads to  substantial improvements in both scale uncertainty of the prediction as well as description
of the data in general. 

While in the $p_T^\gamma$ spectra of inclusive photon events discussed earlier, the NNLO corrections were 
largely flat over the whole range, this is not the case for the inclusive photon-plus-jet process. The corrections are
 negative for $p_T^\gamma < 175~\GeV$ and small and positive for   $p_T^\gamma > 175~\GeV$, 
they change the shape of the distribution, so that it describes that of the data much better,  particularly up to $550~\GeV$. The improvement in the scale uncertainty is similar to what we have 
observed previously for inclusive photon production, going down from approximately $\pm10\%$ at NLO to no more than $(+2.3,-4.7)\%$ at NNLO\@. In most bins the uncertainty is even smaller than that. The NNLO scale band lies within the NLO band, pointing towards convergence of the perturbative series. In the last bin for which data is available, the calculation overestimates the cross section. In this region, electroweak Sudakov logarithms start to become numerically sizable~\cite{Becher:2013zua,Becher:2015yea}, and 
could be resolved with increasingly accurate data. 
\begin{figure}[!t]
    \centering
    \includegraphics[scale=1.2]{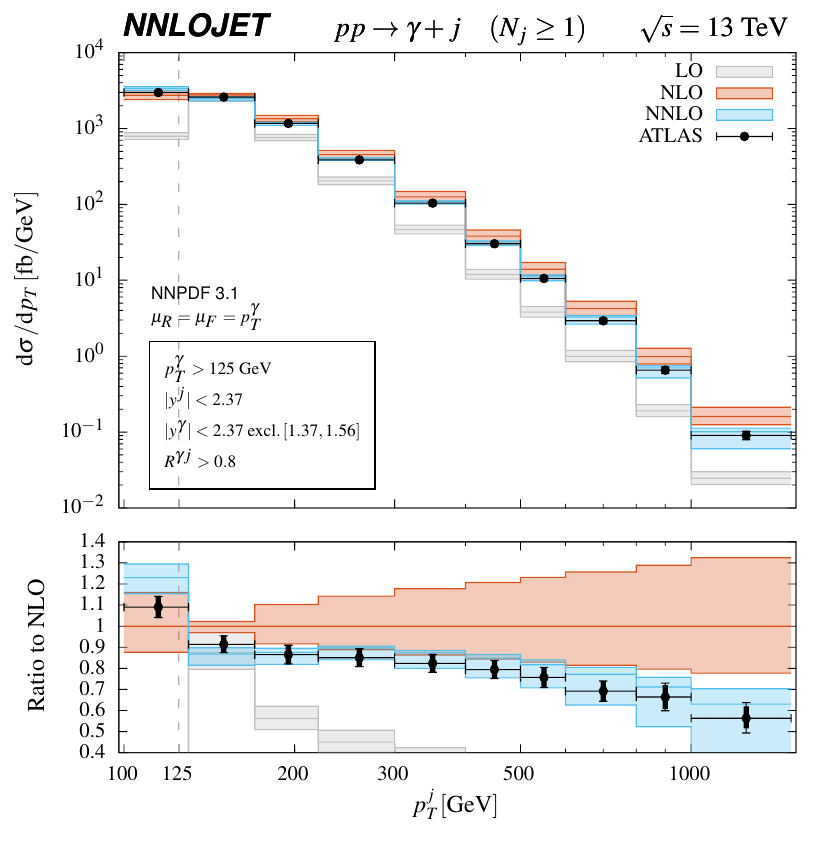}
    \caption{Transverse momentum distribution of the leading jet in photon-plus-jet events, at LO, NLO and NNLO\@. The grey dashed line marks the cut on the $p_T$ of the photon, the recoil of which mostly goes into the leading jet. The predictions are compared to ATLAS data~\protect{\cite{Aaboud:2017kff}}.}%
    \label{fig:GJ_13_A_etj}
\end{figure}

\begin{figure}[!t]
    \centering
    \includegraphics[scale=1.2]{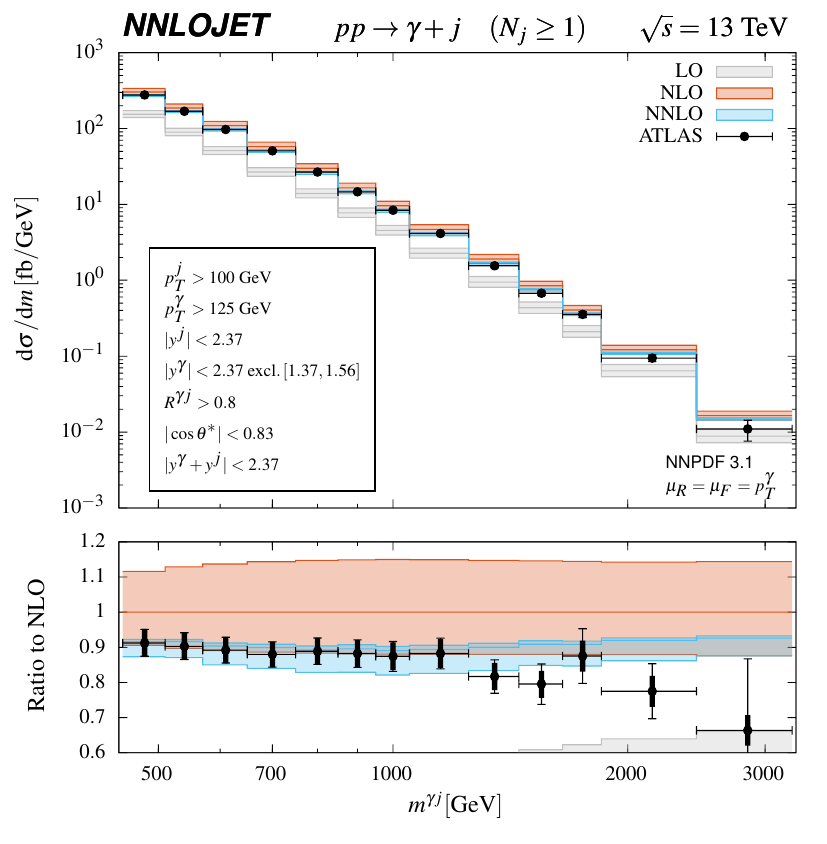}
    \caption{Invariant mass of the photon leading jet system, at LO, NLO and NNLO. The predictions are compared to ATLAS data~\protect{\cite{Aaboud:2017kff}}.}%
    \label{fig:GJ_13_A_mGj}
\end{figure}
At leading order, the photon and the leading jet carry identical amounts of transverse momentum. Including higher order corrections, this one-to-one correspondence no longer holds, although 
typically photon and leading jet largely balance each other's transverse momenta. The leading jet  $p_T^j$ distribution is shown in 
figure~\ref{fig:GJ_13_A_etj}.
The fiducial $p_T$-cuts~\eqref{eq:ATLAScuts13} on the photon and the jets are slightly different, which 
leads to a discontinuity in the LO $p_T^j$ spectrum around the value of the $p_T^\gamma$ cut, marked in the plot with a dashed line. 
Being forced into a strict back-to-back configuration at LO, in all events the jet has at least $125~\GeV$ of transverse momentum, cutting a significant portion 
from of the first bin ($(100-130)~\GeV$). 

As a consequence the cross section in that bin is underestimated quite significantly. Only from NLO onwards, when additional real radiation can 
take part of the recoil, we can have a softer leading jet, thereby describing the event kinematics more truthfully, and leading to a better agreement with the data. 
The cross section in the first bin is therefore described at one order lower than it is in the other bins, which reflects itself in the size of the NLO scale variation, being significantly larger in the first bin  than in the second. 
The NLO corrections and the associated scale uncertainty increase very substantially towards larger $p_T^j$. This effect stems from configurations with two hard back-to-back jets accompanied by  photon at much lower transverse momentum, which is effectively described as a leading order process. 
All but the first data point lie below the NLO uncertainty band, which fails to describe the shape of the data. 
It is only upon including the NNLO corrections that the theory prediction matches the measured spectrum, and 
that scale errors become more uniform at $(+3,-11)$\% size, at least for the $p_T^j$ range from $130~\GeV$ to $500~\GeV$. For higher $p_T^j$ the scale uncertainty grows rapidly, up to $(+11,-40)$\% in the highest bin. The origin of this can be found in the same configurations which already inflated the NLO scale uncertainty at high $p_T^j$. 

The invariant-mass distribution of the photon--jet system is shown in figure~\ref{fig:GJ_13_A_mGj}. As for the $p_T^j$-distribution, we observe very large and positive NLO corrections.    Here the NLO scale uncertainty is again around $(+12,-10)\%$ for the low mass bins and growing moderately to roughly $(+15,-12)\%$ in the bins above $1000~\GeV$. The NNLO correction is nearly constant and shifts the central value towards the lower edge of the NLO scale band, while decreasing the scale uncertainty to no more than $(+1.4,-7.7)\%$. With this the NNLO result matches the data nicely up to $1250~\GeV$. At higher masses the measured cross section lies below the prediction in most bins, yet still being consistent within increasing errors.

\begin{figure}[!t]
    \centering
    \includegraphics[scale=1.2]{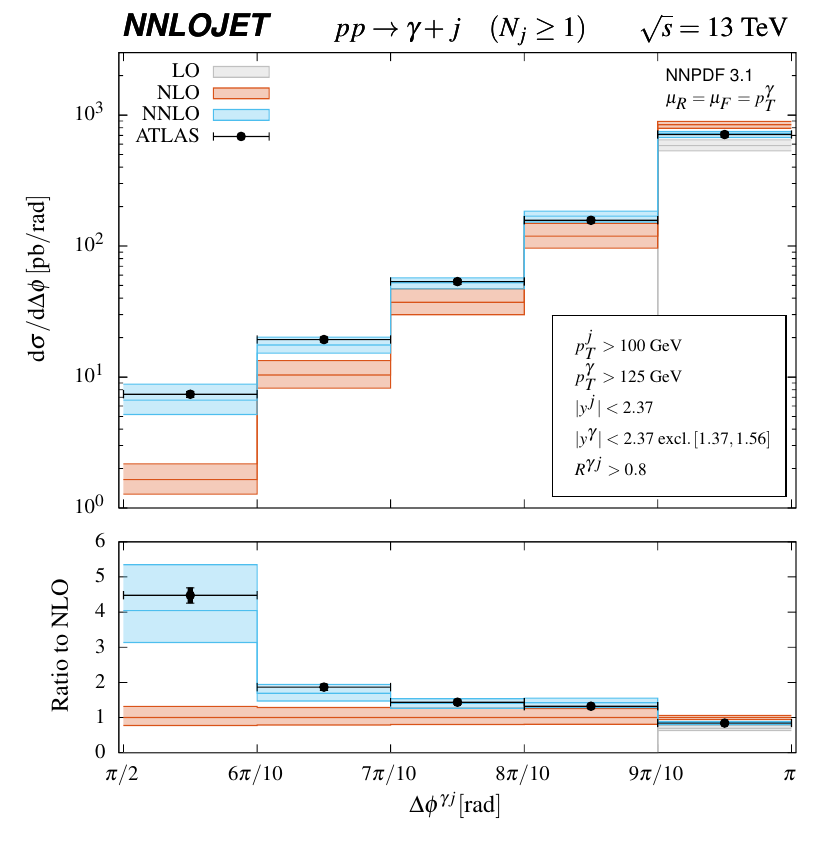}
    \caption{Azimuthal separation of the photon and the leading jet, at LO, NLO and NNLO\@. The predictions are compared to ATLAS data~\protect{\cite{Aaboud:2017kff}}.}%
    \label{fig:GJ_13_A_dpGj}
\end{figure}
The azimuthal separation in the photon--jet system, shown in figure~\ref{fig:GJ_13_A_dpGj}, is described in a meaningful manner only from NLO onwards, as in the LO configuration 
photon and the leading jet (the only jet present in the event) are exactly back-to-back and thus $\left.\Delta\phi^{\gamma j}\right|_\mathrm{LO}\equiv\pi$. It is 
closely related to the azimuthal separation in diphoton production, whose perturbative description has been investigated in detail~\cite{Catani:2011qz,Campbell:2016yrh}. 
The NLO description of this observable is still dominated by back-to-back configurations and fails to provide a decent description of the data~\cite{Aaboud:2017kff}. 
Only after including the NNLO corrections, allowing for one more real radiation parton to take part of the recoil and shifting the leading jet away from the back-to-back configuration, we see a significant enhancement in smaller separation angles. In particular in the lowest bin ($\pi/2$ to $3\pi/5$), the prediction is increased by more than a factor of four compared to its NLO value. The scale uncertainty is of similar size at NLO and NNLO, and in particular for smaller angles. At NNLO it decreases from $(+32,-22)\%$ in the lowest bin to $(+3.0,-7.0)\%$ in the back-to-back bin, this one being effectively one order higher than the others. Still the predictions match the data quite well and it becomes obvious that an NNLO calculation is indeed needed to make reasonable theoretical predictions about this specific observable.  

\begin{figure}[!t]
    \centering
    \includegraphics[scale=1.2]{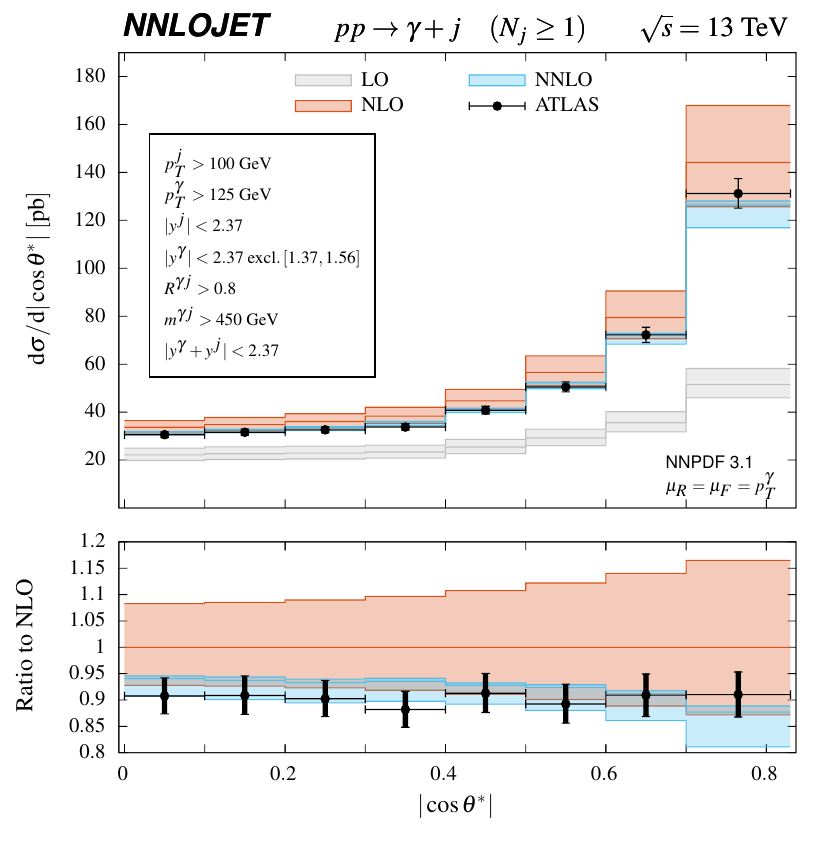}
    \caption{Distribution in $\left|\cos\theta^*\right|$ at LO, NLO and NNLO\@. The predictions are compared to ATLAS data~\protect{\cite{Aaboud:2017kff}}.}%
    \label{fig:GJ_13_A_cts}
\end{figure}
\begin{figure}[!t]
    \centering
    \includegraphics[scale=0.9]{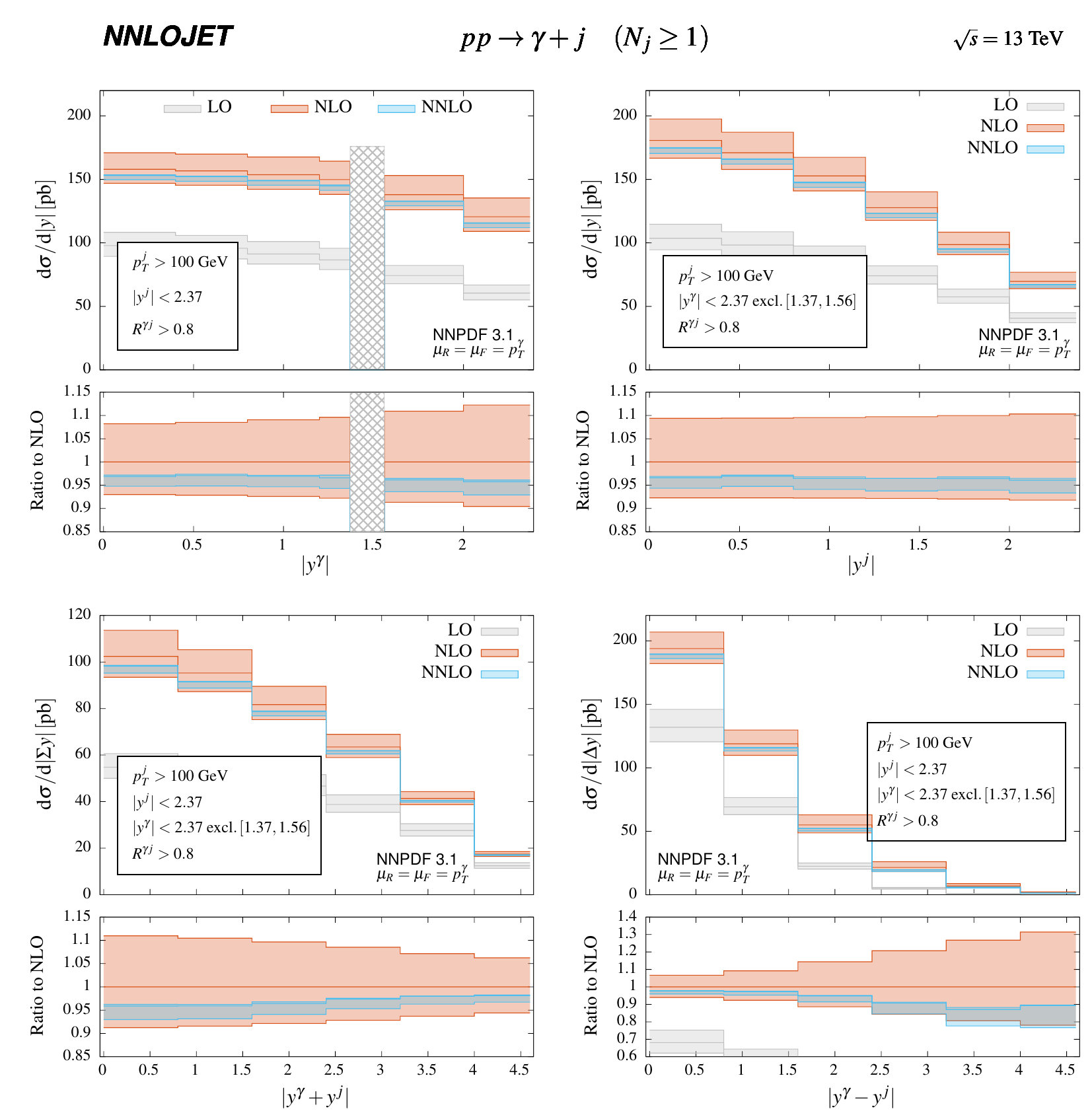}
    \caption{Rapidity distribution of the photon (top left), of the leading jet (top right), distribution of the rapidity sum of both (bottom left) and of the rapidity difference (bottom right), at LO, NLO and NNLO.}%
    \label{fig:GJ_13_A_y}
\end{figure}    
Figure~\ref{fig:GJ_13_A_cts} shows the distribution in $|\cos\theta^*|$, which represents the scattering angle~\eqref{eq:costhetastar} in the underlying two-to-two Born process. On this distribution,
the additional cuts~\eqref{eq:ATLAShighmass} were applied, thereby selecting photon--jet systems with high invariant mass. Its perturbative behaviour is thus similar to the  $p_T^j$ and $m^{\gamma j}$
distributions, with very large and positive NLO corrections.  
We also find the NNLO corrections to be negative, shifting the central prediction to the lower edge of the NLO scale band. The scale uncertainty is reduced significantly at NNLO 
to no more than $(+1.3,-7.5)\%$, in most bins it is even smaller.

This observable was discussed  by ATLAS~\cite{Aaboud:2017kff} in view of a possible sensitivity to the photon fragmentation function at large  $|\cos\theta^*|$, arising from differences in 
the  angular dependence 
of the underlying Born process for direct production and fragmentation. The hybrid isolation procedure used in our calculation eliminates the photon fragmentation contribution, it is however 
only an approximation
to the fixed-cone isolation used in the ATLAS measurement. Given that our calculation provides already  a very good description of the $|\cos\theta^*|$ distribution, being consistent within 
errors throughout the full kinematical range, we conclude that the data at large  $|\cos\theta^*|$ leave only little room for a contribution from photon fragmentation. 
Instead of investigating 
specific kinematical regions in isolated photon production, a more promising approach to the determination of the photon fragmentation functions may be through in the study of 
non-isolated photons inside hadronic jets~\cite{Glover:1993xc,GehrmannDeRidder:2006vn,Kaufmann:2016nux}.

The ATLAS measurement~\cite{Aaboud:2017kff} of photon-plus-jet production was performed inclusively in rapidity. To gain better insight in the kinematical dependence of the perturbative corrections, and 
in the potential sensitivity to the parton distributions, figure~\ref{fig:GJ_13_A_y} displays the rapidity distributions of the photon and the leading jet, as well as distributions in rapidity sum and difference. 
While the NNLO corrections appear to be quite uniform in the individual rapidity distributions, we observe some changes of the shapes in the rapidity correlation distributions. In these, the 
corrections are largest in magnitude for small rapidity sum (symmetric events) and for large rapidity difference (small scattering angle), remaining negative throughout. 

\subsection{Comparison with CMS 13 TeV measurements}%
\label{subsec:GJ_13_C}

\begin{figure}[!t]
    \centering
    \includegraphics[scale=0.9]{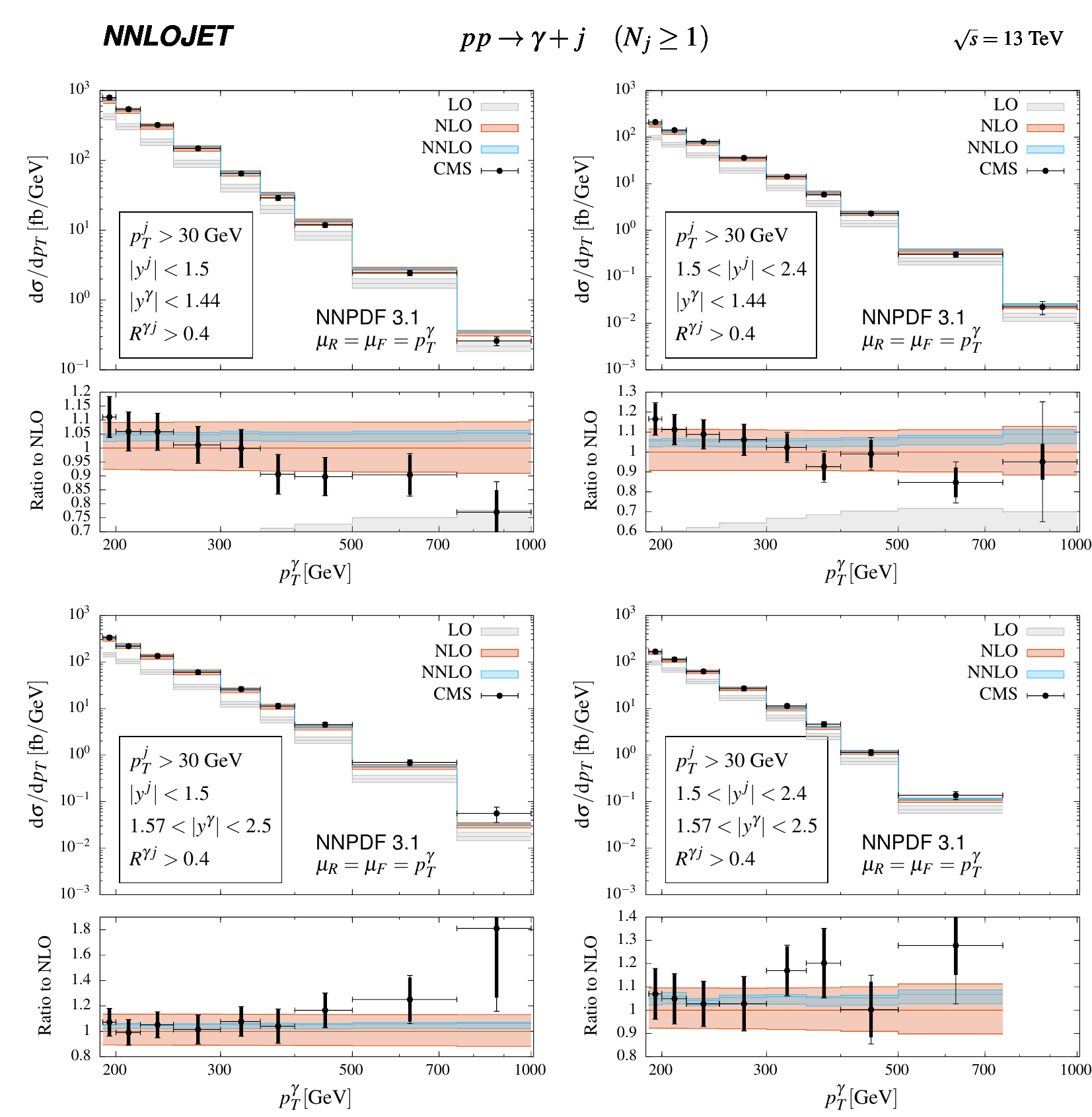}
    \caption{Transverse energy/momentum distribution of the photon at LO, NLO and NNLO in two different rapidity bins for the photon and the leading jet, each. The results are compared to CMS data~\protect{\cite{Sirunyan:2018gro}}. Note that the data has been multiplied by the corresponding rapidity bin-widths, as CMS presents the data in triple-differential form in ($p_T^\gamma,y^\gamma,y^j$).}%
    \label{fig:GJ_13_C_etG}
\end{figure}
The $13$~TeV CMS study of isolated photon production~\cite{Sirunyan:2018gro} discussed in Section~\ref{subsec:iG_13_C} above also provides measurements of photon-plus-jet 
observables. For these, jets are clustered using the anti-$k_T$ algorithm with radius parameter $R^j=0.4$ and requiring 
\begin{align}
    p_T^j &> 30~\GeV\,,&  p_T^\gamma &> 190~\GeV\,, &  R^{\gamma j} &> 0.4\,.
\end{align}
Results for the photon transverse momentum distribution are presented in different bins in rapidity for the photon and the jet, corresponding to central and forward production:
\begin{align}
    |y^\gamma| &\in [0,1.44]    & \mathrm{and}  &&  |y^\gamma| &\in [1.57,2.5]\,,\\
    |y^{j}| &\in [0,1.5]    & \mathrm{and}  &&  |y^{j}| &\in [1.5,2.4]\,,
\end{align}          
omitting the region $|y^\gamma| \in [1.44,1.57]$. With these, four combinations of central/forward  photon/jet rapidity regions are measured. 

We use our default setup, with NNPDF3.1 and hybrid isolation using the same parameters~\eqref{eq:CMSiso} as in Section~\ref{subsec:iG_13_C} and  $\mu_R=\mu_F=p_T^\gamma$  as 
central scale choice.
The results are shown in figure~\ref{fig:GJ_13_C_etG}.
We find the NLO scale uncertainty to be largely flat over the whole $p_T^\gamma$ range in all four rapidity bins, with a size of about $\pm 10\%$. The NNLO corrections are positive and mostly flat in all bins, increasing the central prediction by $(4-9)\%$. The scale uncertainty is reduced to less than $(+0.9,-2.8)\%$ for the configurations with a central jet and  less than $(+2.1,-4.3)\%$ for those with a forward jet. 

Although the predictions are consistent with the data within the respective uncertainties, they do however not yield a good description of the shape of the measurement. 
It appears that the form of the discrepancy depends on the photon rapidity and not so much on the leading jet rapidity. For the central photon the prediction is too low in the low $p_T^\gamma$ region and too high in the high $p_T^\gamma$ region, irrespectively of the jet rapidity. If the photon is forward, the predictions match the data at low $p_T^\gamma$, but underestimate the cross section in the high $p_T^\gamma$ tail, again independently of the jet rapidity. In other words, the calculation predicts more high-$p_T$ photons in the central region than there are actually observed, but fewer softer photons in the central and fewer hard photons in the forward region. 
A similar pattern was already observed in the isolated photon distribution measured by CMS, figure~\ref{fig:iG_13_C_etG}, although somewhat less pronounced. 

If this tension in the shape persists and becomes more pronounced, it will be rather unlikely that it could be accommodated by a modification of the parton distribution functions. In the case of the photon-plus-jet 
measurement, it may indicate that the rather low jet transverse momentum cut leads to potentially large logarithmic corrections, which are poorly described by fixed-order perturbation theory.

\section{Summary and Conclusions}%
\label{sec:conc}

Isolated photon and photon+jet production constitute essential probes to test perturbative QCD predictions and can provide important constraints on the gluon distribution inside the proton.
The detailed study of these processes, however, requires isolation cuts on the photon in order to suppress the overwhelming background of secondary photons, e.g.\ coming from the decay of $\pi^0$. Since many years, a disparity persists between the isolation prescriptions used in theory predictions and the experimental measurement:
While all measurements are performed using a \emph{fixed cone isolation}, theory calculations are commonly performed using a \emph{dynamical cone isolation} in order to avoid the complications that 
the fragmentation component of the process entails (in particular the dependence on the non-perturbative part of the photon fragmentation function, which is only loosely constrained 
by experimental measurements). 
This mismatch has been the subject of many studies that concluded with recommended settings for the isolation that aim to reduce its numerical impact, which however can still be at the level of a few per cent.
With the advent of NNLO calculations and the increasingly more precise measurements performed at the LHC, per-cent-level studies of this process are now a reality and the impact of this mismatch needs to be revisited and ideally improved on.

A first step towards bringing the theory predictions closer in line with experiment is given by the \emph{hybrid cone isolation}, which embeds a smooth cone isolation prescription with a narrow cone $R_d$ within the standard isolation with a fixed cone $R$. 
We performed a detailed study of isolation settings at NLO, which revealed only a moderate sensitivity on the inner cone's parameters of the hybrid prescription, allowing us to infer a much reduced 
ambiguity associated with this procedure.
In the limit of small $R_d$, we further confirmed the correct logarithmic behaviour $\sigma\sim\log(1/R_d)$ as predicted by QCD. This behaviour is 
compensated in the limit $R_d\to 0$ by a negative counter-term from the photon fragmentation function, resulting in a numerical compensation of 
fragmentation contributions and dynamical cone suppression at a finite value of $R_d$, which we estimate to be in the vicinity of $R_d=0.1$. 
The associated uncertainty on the cross section predictions is at the level of a few per cent, and largely concentrated at low values of photon transverse momentum. 
For $p_T^\gamma>125$~GeV, a very conservative estimate based on comparison at NLO with fixed-cone isolation and a model for the photon fragmentation functions 
results in below 5\% uncertainty, while a variation of the $R_d$ parameter at NNLO points to an uncertainty below 2\%. 
Moreover, using the hybrid isolation the exact dependence of the cross section on the fixed cone $R$ can be predicted, as long as $R_d^2 \ll R^2$ is respected.
This opens up the possibility to perform more stringent tests of perturbative QCD predictions in the future once measurements are available for different cone sizes.

Up to now, all predictions at NNLO in QCD involving prompt photons have employed a dynamic cone isolation.
In this paper we presented our calculation of isolated photon and photon+jet processes at NNLO in QCD using the antenna subtraction method and, for the first time, apply the hybrid cone isolation at this order.
We performed a detailed comparison of our predictions to the available measurements by the ATLAS and CMS collaborations, which overall show an excellent agreement to the data.
Going from NLO to NNLO, we observe a dramatic reduction in scale uncertainties across the entire kinematic range with residual scale uncertainties that are typically at the level of $5\%$ or smaller for genuine NNLO observables.
For isolated photon production, the NNLO corrections are rather flat with a positive shift of about $+5\%$ in the central prediction. 
In the photon-plus-jet process, on the other hand, NNLO corrections can induce substantial shape distortions that often could not be resolved at NLO due to the much larger scale uncertainties. 
The reduced theory errors further expose some minor tensions with the CMS photon+jet measurement at 13~TeV, which are difficult to account for by PDF effects and will require further investigation. 

The excellent perturbative convergence displayed in the NNLO prediction combined with a photon isolation treatment that follows closely the  procedure used in experiments
puts the theory predictions on a solid basis with residual uncertainty estimates that are competitive with the experimental errors, often even surpassing them. 
Although much smaller than in the dynamic cone isolation, the hybrid approach still contains an intrinsic 
theoretical ambiguity from the removal of the fragmentation component through the narrow inner cone.
Further progress in alleviating the mismatch between experiment and theory for the isolation procedure will require the calculation of the fragmentation component at NNLO\@. With this,
the theory calculation can apply identical photon isolation criteria as used in the experimental measurements, however at the expense of introducing a novel dependence on 
 photon fragmentation functions.

\acknowledgments 
The authors thank Juan Cruz-Martinez, Rhorry Gauld, Aude Gehrmann-De Ridder, Imre Majer, Jan Niehues, Duncan Walker and James Whitehead for useful discussions and their many contributions to the \nnlojet code. 
The authors also thank the University of Zurich S3IT and CSCS Lugano for providing the computational resources for this project. 
This research was supported in part by the UK Science and Technology Facilities Council under contract ST/G000905/1, by the Swiss National Science Foundation (SNF) under contract 200020-175595,  by the Swiss National Supercomputing Centre (CSCS) under project ID UZH10, and by the Research Executive Agency (REA) of the European Union through the ERC Consolidator Grant HICCUP (614577) and the ERC Advanced Grant MC@NNLO (340983).

\bibliographystyle{JHEP}
\bibliography{photon_rev}


\end{document}